%% file: CP2Vortices.tex
\documentclass[11pt,twoside]{article} 

\usepackage{amsmath}
\usepackage{latexsym,amssymb}
\usepackage{amsthm}
\usepackage{hyperref}
\usepackage{titlesec} 
\usepackage{titling} 

\def\titlecolour{\color{DarkSlateBlue}}
\def\seccolour{\color{Crimson}}
\def\headcolour{\color{DarkGrey}}

\pretitle{\begin{center}\LARGE\bfseries\sffamily\titlecolour}
\preauthor{\begin{center}\Large\sffamily\titlecolour}
\postauthor{\par\end{center}}

\titleformat{\section}{\Large\bfseries\sffamily\seccolour}{\thesection}{1em}{}
\titleformat{\subsection}{\large\bfseries\sffamily\seccolour}{\thesubsection}{1em}{}
\titleformat{\paragraph}[runin]{\bfseries\sffamily\seccolour}{}{0.5em}{}
\titleformat{\subsubsection}[runin]{\small\sffamily\seccolour}{}{0.5em}{}

\usepackage[utf8]{inputenc} 
\usepackage[margin=1cm,font=small,labelfont=bf]{caption}
\usepackage{subcaption}

\usepackage{geometry} 
\usepackage{graphicx} %
\usepackage[svgnames]{xcolor}                    
\usepackage{tikz,tikz-cd}

\usepackage{euler}                             
\usepackage{times}

\usepackage{float}
\usepackage{enumerate}

\usepackage{booktabs} 
\usepackage{multirow,bigdelim,colortbl} %
\usepackage{array} 

\usepackage{fancyhdr}
\usepackage[nottoc,notlof,notlot]{tocbibind} 
\usepackage[titles,subfigure]{tocloft} 

\usepackage[T1]{fontenc}

\hypersetup{colorlinks=true,linkcolor=DarkSlateBlue,citecolor=DarkSlateBlue}

\DeclareMathAlphabet{\mathcal}{OMS}{cmsy}{m}{n}

\numberwithin{equation}{section}
\numberwithin{figure}{section}
\numberwithin{table}{section}

\newtheoremstyle{theoremsf}
{2ex}
{2ex}
{\itshape}
{}
{\sffamily\seccolour}
{}
{1em}
{}

\newtheoremstyle{definitionsf}
{2ex}
{2ex}
{}
{}
{\sffamily\seccolour}
{}
{1em}
{}

\theoremstyle{theoremsf}
\newtheorem{theorem}{Theorem}[section]
\newtheorem{lemma}[theorem]{Lemma}
\newtheorem{proposition}[theorem]{Proposition}

\theoremstyle{definitionsf}
 \newtheorem{definition}[theorem]{Definition}
 
 \newtheorem{remark}[theorem]{Remark}

\def\gg{\mathfrak{g}}
\def\so{\mathfrak{so}}
\def\su{\mathfrak{su}}

\def\nn{\mathfrak{n}}
\def\tt{\mathfrak{t}}

\def\u{\mathfrak{u}}

\def\TT{\mathbb{T}}

\def\R{\mathbb{R}}
\def\C{\mathbb{C}}

\def\J{J} 
\def\j{\mathcal{J}}
\def\cO{\mathcal{O}}
\def\CP{\mathbb{CP}}

\def\cP{\mathcal{P}}

\def\diag{\mathop\mathrm{diag}\nolimits}
\def\Fix{\mathop\mathrm{Fix}\nolimits}

\def\IntT{\mathrm{Int}(\tt^*_+)}

\def\e{\mathsf{e}}
\def\i{\mathsf{i}}

\def\defn#1{{\bfseries\itshape #1}}

\def\restr#1{\vrule height1.2ex width.4pt
  depth1.4ex\lower0.8ex\hbox{\scriptsize $\,#1$}}

\def\Ma{M_{(a)}} 
\def\Mb{M_{(b)}} 
\def\Ja{\J_{(a)}} 
\def\Jb{\J_{(b)}} 

\newcommand{\WeylChamber}{%
\path [fill=Pink!50] (0,0) -- (0,10) .. controls (3,9) and  (6,8) .. (8.66,5) -- (0,0);
\draw [blue] (0,0) -- (0,10);
\draw [blue] (8.66,5) -- (0,0);
}

\newcommand\blfootnote[1]{%
  \begingroup
  \renewcommand\thefootnote{}\footnote{#1}%
  \addtocounter{footnote}{-1}%
  \endgroup
}

\title{Generalized point vortex dynamics on $\CP ^2$}
\author{James Montaldi \& Amna Shaddad \\[6pt]\small University of Manchester}
\date{}

\begin{document} 
\maketitle
\thispagestyle{empty}

\vbox to 0pt{\hfill\emph{Dedicated to Darryl Holm on the occasion of his 70th birthday}\par\vss}

{\small
\noindent\hrulefill 

\smallskip

\noindent{\large\sffamily\color{DarkSlateBlue} Abstract}

\medskip

\noindent 
This is the second of two companion papers. 
We describe a generalization of the point vortex system on surfaces to a Hamiltonian dynamical system consisting of two or three points on complex projective space $\CP ^2$ interacting via a  Hamiltonian function depending only on the distance between the points.  The system has symmetry group SU(3). The first paper describes all possible momentum values for such systems,  and here we apply methods of symplectic reduction and geometric mechanics to analyze the possible relative equilibria of such interacting generalized vortices. 

The different types of polytope depend on the values of the `vortex strengths', which are manifested as coefficients of the symplectic forms on the copies of $\CP ^2$.  We show that the reduced space for this Hamiltonian action for 3 vortices is generically a 2-sphere, and proceed to describe the reduced dynamics under simple hypotheses on the type of Hamiltonian interaction.  The other non-trivial reduced spaces are topological spheres with isolated singular points. 
For 2 generalized vortices, the reduced spaces are just points, and the motion is governed by a collective Hamiltonian, whereas for 3 the reduced spaces are of dimension at most 2. In both cases  the system will be completely integrable in the non-abelian sense.

\medskip


\noindent\hrulefill 
}

\bigskip
\begin{center}
\begin{tikzpicture}[scale=0.35] 
	\draw[very thick,fill=Purple] (0, 0) -- (4.157, 7.2) -- (1.039, 7.200) --
	 (0, 5.4) -- (-1.039, 7.200) -- (-4.157, 7.2) -- (0,0);
	\draw[very thick,fill=Purple,rotate=120] (0, 0) -- (4.157, 7.2) -- (1.039, 7.200) --
	 (0, 5.4) -- (-1.039, 7.200) -- (-4.157, 7.2) -- (0,0);
	\draw[very thick,fill=Purple,rotate=-120] (0, 0) -- (4.157, 7.2) -- (1.039, 7.200) --
	 (0, 5.4) -- (-1.039, 7.200) -- (-4.157, 7.2) -- (0,0);
	\draw[thin,blue] (0,-7) -- (0,9);
	\draw[thin,blue,rotate=120] (0,-7) -- (0,9);
	\draw[thin,blue,rotate=-120] (0,-7) -- (0,9);
 \end{tikzpicture}
\end{center}

\newpage

\tableofcontents

\blfootnote{The picture on the front cover is the intersection of the image of the momentum map with the dual of the Cartan subalgebra, for vortex strengths (symplectic weights) of type D$_0$ (see \cite[Fig.\,4.7]{MS19a}): it does not appear in the published version. }

\section{Introduction}

A point vortex is a point of isolated vorticity traditionally in a background of irrotational fluid, although in some cases one now allows a background of constant vorticity (necessary for arbitrary point vortices on a compact surface).  It was discovered by Helmoltz in 1858 that point vortices interact and evolve in a way that depends only on their strengths and mutual positions. Some years later (1876) Kirchhoff noticed that the equations of motion could be written in Hamiltonian form---perhaps the first example of a Hamiltonian system not governed by kinetic and potential energy. It was through the work of Novikov in the 1970s that the Hamiltonian formulation of point vortex dynamics came to the fore, and this approach has produced many outstanding advances.  Among these are the complete integrability of the motion of three vortices on a sphere proved by Kidambi \& Newton \cite{KN98}, the existence of quasiperiodic orbits on invariant tori for lattice vortex systems demonstrated by Lim \cite{Lim90}, and several other related results.  In more recent years, the subject has been extended to the study of  point vortices on the sphere \cite{PM98,L-PMR,LMR}, on the hyperbolic plane \cite{MN-G}, and other less symmetric surfaces \cite{BK,RCK}.  The study of point vortices has been described as a mathematics playground by H.~Aref  \cite{Aref-playground}, by which he meant that many different areas of (classical) mathematics can be brought to bear to study these point vortex systems. 

The general Hamiltonian set-up is as follows.  Let $(S,\omega_0)$ be a symplectic surface (smooth manifold of dimension 2, often endowed with a Riemannian metric).  A configuration is an ordered set of $n$ distinct points in $S$, and hence an element of
$$M=S\times S\times \dots \times S \; \setminus \; \Delta.$$
Here $\Delta$ is the `large diagonal', the subset consisting of all possible collisions: it is customary to rule collisions out of the model. 
Each of the $n$ points $x_j\in S$ has a fixed non-zero real number $\Gamma_j$ associated to it, called the  \emph{vortex strength}.  The symplectic form $\Omega$ on $M$ is defined to be 
\begin{equation}\label{eq:symplectic form}
\Omega=\Gamma_1\omega_0\oplus\Gamma_2\omega_0 \oplus\cdots\oplus \Gamma_n\omega_0,
\end{equation}
and indeed the $\Gamma_j$ are called the \emph{symplectic weights} in \cite{MS19a}. 
More formally, if $\pi_j:M\to S$ is the Cartesian projection to the $j^{\mathrm{th}}$ component of $M$, then
$$\Omega=\Gamma_1\pi_1^*(\omega_0)+\Gamma_2\pi_2^*(\omega_0)+\cdots+\Gamma_n\pi_n^*(\omega_0).$$
For this to be a symplectic form one requires that all $\Gamma_j\neq0$; we make this assumption throughout.

The dynamics of a point vortex system is defined by a pairwise interaction: let $h_0:S\times S\setminus\Delta\to\R$ be a given  smooth function (usually taken to be the negative of the Green's function of the Laplacian relative to the Riemannian metric).  The Hamiltonian function $H:M\to\R$ is then\footnote{More generally \cite{DB2015} there may be a `self-interaction' term $\sum_j\Gamma_j^2R(x_j)$ for some function $R$ (known as the Robin function), encoding the interaction of the point vortex with the geometry or asymmetry of the space.  We ignore this as $\CP^2$ is highly symmetric. 
} 
$$H(x_1,\dots,x_n) = \sum_{i<j}\Gamma_i\,\Gamma_j \,h_0(x_i,x_j),$$
The evolution of the system is given by, 
$$\dot x = X_H(x)$$
where the vector field $X_H$ is defined by Hamilton's equation $i_{X_H}\Omega = -dH$. 

If in addition there is a group $G$ acting on the surface $S$ preserving the symplectic form, and preserving the function $h_0$ (that is $h_0(g\cdot x,g\cdot y) = h_0(x,y)$ for all $g\in G, \; x,y\in S$), then the point vortex system has symmetry $G$ and the vector field is equivariant.  Moreover, if the action on $(S,\omega_0)$ is \emph{Hamiltonian}, meaning that there is a momentum map $\J_0:S\to\gg^*$, then so is the action on $(M,\Omega)$ with momentum map $\J:M\to\gg^*$ given by 
\begin{equation}\label{eq:general mommap}
\J(x_1,\dots,x_n) = \sum_j \Gamma_j\,\J_0(x_j).
\end{equation}
This approach highlights the application of Noether's theorem, which states that the components of the momentum map are preserved by the dynamics. 

If $S$ is the plane, the sphere or the hyperbolic plane then the respective actions of the groups $SE(2)$, $SO(3)$ and $SL(2)$ are indeed Hamiltonian (see above references and \cite{JM-peyresq98}). However, if the point vortices lie on a torus or cylinder then the actions of $\TT^2$ and $\R\times S^1$ respectively, while being symplectic, fail to be Hamiltonian \cite{MST}. See also the recent paper about vortices on the round torus \cite{Sakajo-Shimizu}.

In this paper, we extend the playground to higher dimensional symplectic manifolds $(S,\omega_0)$ and in particular to $\mathbb{CP}^2$, and we apply systematic geometric methods to the analysis of relative equilibria for such a system.  We do not assume a particular form for the pairwise interaction, although it would be reasonable to choose the Green's function of the Laplacian as in 2-dimensional point vortices; the only assumption we make is that $h_0(x,y)$ depends only on the distance between $x$ and $y$.  The set-up is otherwise identical to the description above.  It should be emphasized that we are not claiming this model is related to vortex dynamics for 4-dimensional fluids (indeed, it is not clear that isolated points of vorticity can exist in 4 dimensions). 

Previous work relating point vortices to dynamics on $\CP^n$ can be found in Bolsinov, Borisov and Mamaev \cite{BBM}.  Their work reduces the dynamics of $n$ point vortices with positive vorticity in the plane to a dynamical system on $\CP^{n-2}$ (by a process which is essentially symplectic reduction).  In particular the case of 4 vortices in the plane reduces to a dynamical system on (a single copy of) $\CP^2$.  In contrast, we consider (generalized) point vortices on $\CP^2$, which defines a dynamical system on Cartesian products of $\CP^2$.  

The paper is organized as follows.  In order to perform symplectic reduction for this system of generalized point vortices, it is important to know the possible values of the momentum map and its singularities.  This was carried out in the companion paper \cite{MS19a}.  In Section\,\ref{sec:reduction} we discuss the reduced spaces; for 2 vortices these are just single points, while for 3 they are usually diffeomorphic to a sphere, sometimes to a sphere with singular points and for some points on the boundary of the polytope they reduce to a single point; this is made precise in Theorem\,\ref{thm:reduction N=3}; the section ends with a remark on the Duistermaat-Heckman theorem.  Finally, in Section\,\ref{sec:dynamics} we consider the resulting reduced dynamics and in particular the reduced and relative equilibria and their stability where possible, without reference to the specific form of the Hamiltonian, beyond its symmetry.  For two generalized vortices, every configuration is a (stable) relative equilibrium, but for three this is not the case in general, but is when the reduced space is a point.

Much of this work forms part of the PhD thesis of the second author \cite{AS-thesis}, where further details and alternatives for some of the calculations may be found.

\section{Symplectic reduction}
\label{sec:reduction}
When studying the dynamics of symmetric Hamiltonian systems, a first approach is to study the reduced systems.  Recall that the symplectic reduction at $\mu\in\gg^*$ is the space
$$M_\mu := \J^{-1}(\mu)/G_\mu \simeq \J^{-1}(\cO_\mu)/G,$$
where these two versions of the reduced space are known as point reduction and orbit reduction. A proof of the fact that they are equivalent for compact groups can be found in \cite[Theorems 6.4.1 \& 8.4.4]{OrteRati04}.  If $\mu$ is a regular (ie, non-singular) value of the momentum map, 
or equivalently, at all points of the fibre $\J^{-1}(\mu)$ the group action is (locally) free,  then by a dimension count the reduced spaces are smooth manifolds of dimension
\begin{equation}\label{eq:dim of reduced space}
\dim M_\mu = \dim M - \dim G -\dim G_\mu.
\end{equation}

One advantage of orbit reduction is that it allows for studying the variation of reduced spaces as the momentum value varies.  Indeed, since the momentum map is equivariant, it descends to a map between orbit spaces we call the \emph{orbit momentum map} and denote $\j$, according to the following diagram,
\begin{equation}\label{eq:OMM}
  \begin{tikzcd}
     M \arrow[r,"\J"]\arrow[d] &\su(3)^*\arrow[d]\\
     M/G\arrow[r,"\j"] &\gg^*/G 
  \end{tikzcd}
\end{equation}
where the vertical maps are the quotient maps. The fibres of $\j$ are the reduced spaces, using orbit reduction. 

The momentum map on the phase space of interest, namely products of $\CP^2$, is determined by \eqref{eq:general mommap}, where $\J_0:\CP^2\to\su(3)^*$ is given by
$\J_0(Z) = Z\otimes\overline{Z}-\frac13I_3$ (see \cite{MS19a}).  Explicitly,
$$\J_0([x:y:z]) = \begin{pmatrix}
|x|^2-\frac13&x\bar y & x\bar z \cr \bar x y & |y|^2-\frac13 & y\bar z\cr \bar x z & \bar y z &|z|^2-\frac13
\end{pmatrix}.
$$
Here we consider $\CP^2$ as $S^5/U(1)$, so that $|x|^2+|y|^2+|z|^2=1$, and the matrix has trace 0 as required. 

The distance between two points in $\CP^2$ can be given by the dihedral angle between the complex lines represented by the points. A formula is, 
\begin{equation}\label{eq:metric}
d(Z_1,\,Z_2) = \arccos\left| \widehat Z_1^\dagger \,\widehat Z_2 \right|.
\end{equation}
Here $\widehat Z$ is any representative of $Z$ in $S^5$, and $W^\dagger=\overline{W}^T$ is the conjugate transpose, so the expression within the modulus is the complex inner product.  Note that this expression is well-defined: $|W_1^\dagger\,W_2|=\left|(\e^{\i\theta}W_1)^\dagger\,(\e^{\i\phi}W_2)\right|$.  

The minimal distance between two points in $\CP^2$ is of course 0, and the maximum is $\pi/2$, when the points are orthogonal\footnote{note that under the correspondence of $\CP^1$ with $S^2$, points are orthogonal in $\CP^1$ iff the corresponding points in $S^2$ are antipodal}.   Given any two points separated by a distance $\theta$, there  is an element of $SU(3)$ that transforms them into $e_1$ and $\cos\theta e_1+\sin\theta e_2$ (where linear combination is understood in $\C^3$, before taking the quotient to $\CP^2$).  Here and throughout we denote,
$$e_1=[1:0:0],\quad e_2=[0:1:0],\quad e_3=[0:0:1].$$  

There are three particular subgroups of $SU(3)$ that we will refer to frequently: 
\begin{equation}\label{eq:subgroups}
\begin{array}{rcl}
U(2) &=& \left\{\begin{pmatrix}
A&0\cr 0&(\det A)^{-1}
\end{pmatrix} \;\mid A\in U(2)\right\}, \\[10pt]
\TT^2 &=& \left\{\diag[\e^{\i\theta}, \e^{\i\phi}, \e^{-\i(\theta+\phi)}] \;\mid \theta,\phi\in[0,2\pi]\right\}, \rule[-12pt]{0pt}{30pt}\\
U(1) &=& \left\{\diag[\e^{\i\theta}, \e^{\i\theta}, \e^{-2\i\theta}] \;\mid \theta\in[0,2\pi]\right\}.
\end{array}
\end{equation}
The importance of these groups is that the stabilizer  of any point in $\CP^2$ is conjugate to $U(2)$, of any point in $\CP^2\times\CP^2$ is conjugate to one of these three subgroups, and of any point in $\CP^2\times\CP^2\times\CP^2$ is conjugate to one of these three or is trivial. 
 
The fixed point subspaces in $\CP^2$ of these subgroups are,
\begin{equation}\label{eq:fixed point spaces}
\begin{array}{rcl}
\Fix(U(2),\CP^2) &=& \{e_3\}, \\[4pt]
\Fix(\TT^2,\CP^2) &=& \{e_1,e_2,e_3\},\\[4pt]
\Fix(U(1),\CP^2) &=& \left\{[x:y:0] \,\middle|\, [x:y]\in\CP^1\right\} \cup \{e_3\}.
\end{array}
\end{equation}
Note that the first two consist of isolated points, while the third is the disjoint union of a submanifold of dimension 2 and an isolated point.  With the subgroups given, it is straightforward to see that $U(2)$ is the normalizer of the subgroup $U(1)$, that $U(1)$ is the centre of $U(2)$ and that $\TT^2$ is a maximal torus of both $SU(3)$ and $U(2)$.

For the action of $SU(3)$ on the product of any number of copies of $\CP^2$, the image of the momentum map is invariant under the coadjoint action, and so is determined by its intersection with a positive Weyl chamber.  By a theorem of Guillemin and Sternberg (and generalized by Kirwan), this intersection is a convex polytope, the \emph{momentum polytope}, which we denote $\Delta(M)$, where $M$ is the symplectic manifold under consideration. This polytope can be identified with the image of the orbit momentum map $\j$ defined above, as described in \cite{MS19a}. 

\subsection{Reduction for 2 copies of \texorpdfstring{$\CP^2$}{CP2}}

Let $M=\CP^2\times\CP^2$.  Recall from \cite[Sec.\,3]{MS19a} that the momentum polytope $\Delta(M)$ for the $SU(3)$ action on $M$ is a line segment. One endpoint of the segment lies on a wall of the positive Weyl chamber and arises as the momentum value for points on the diagonal in $M$ (distance 0), while the other endpoint is the image of orthogonal pairs of points (distance $\pi/2$) and may or may not lie on a wall, depending on the vortex strengths; see \cite[Figures 3.1 \& 3.2]{MS19a}.

\begin{theorem}\label{thm:reduction 2}
If\/ $\mu\in\Delta(M)$ then the reduced space $M_\mu$ is a single point.
\end{theorem}

\begin{proof}
Since the orbit space $M/SU(3)$ is 1-dimensional (parametrized by the distance $\theta$ as pointed out above), the orbit momentum map is a map between two 1-dimensional manifolds with boundary, and is injective as the only singular points are the endpoints.  The fibres, which coincide with the reduced spaces, therefore consist of a single point. 
\end{proof}

\subsection{Reduction for  3 copies of \texorpdfstring{$\CP^2$}{CP2}}
Now let $M=\CP^2\times\CP^2\times\CP^2$, with symplectic form given by \eqref{eq:symplectic form}, and non-zero vortex strengths $\Gamma_j$. 
We claim that there are only a few types of reduced space for the $SU(3)$ action on $M$. For most values of the momentum, the reduced space is diffeomorphic to a smooth 2-sphere $S^2$.  For exceptional values of momentum, and for generic $\Gamma_j$ it is either diffeomorphic to a \defn{pointed sphere}, which is a topological $S^2$ with a single conical singular point, or equal to a single point.  For special `transition' value of the vortex strengths, there are the additional possibilities of twice- or thrice-pointed spheres.

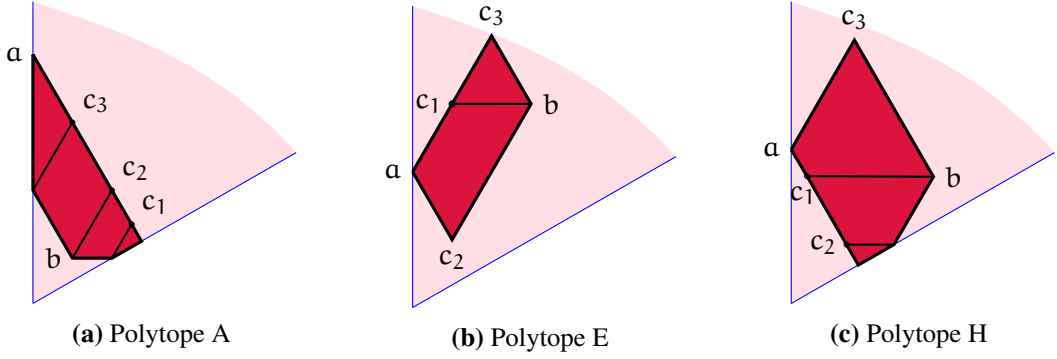
\begin{figure}
	\centering
\input{selectedPolytopes.tex}
\caption{Three of the generic momentum polytopes from \cite{MS19a}. They illustrate in particular how the points $a$ and $b$ are always vertices, but the $c_j$ may or may not be vertices of the polytope.}
\label{fig:polytopes}
\end{figure}

\begin{figure}\label{eq:table of ISGs N=3}
\centering
\begin{subfigure}[b]{0.5\textwidth}
$$
  \begin{tikzcd}[column sep=0.3em]
     &&\mathbf{1} \arrow[dr] \arrow[dl] &&\\
     &U(1)_{(a)} \arrow[dr]\arrow[dl] && U(1)_{(b)}\arrow[dr]\arrow[dl] \\
     U(2) && \TT^2_{(c)} && \TT^2_{(b)} 
  \end{tikzcd}
$$
\caption{Adjacencies of the different orbit type strata:\newline  $X\to Y$ means $Y\subset\overline{X}$. }
\label{fig:adjacencies}
\end{subfigure}
	\hfil
\begin{subfigure}[b]{0.4\textwidth}
\begin{tabular}{c|c}
geometry & stabilizer \\
\hline
triple point & $U(2)$ \\
double+orthogonal & $\TT^2$ --- (c) \\
totally orthogonal & $\TT^2$ --- (b) \\
coplanar & $U(1)$ --- (a) \\
semi-orthogonal & $U(1)$ --- (b) \\
general position & $\mathbf{1}$
\end{tabular}
\caption{Stabilizers for 3 points in $\CP^2$. \newline \ 
}
\label{table:stabilizers}
\end{subfigure}

\bigskip

\caption{This figure and associated table show the orbit type stratification of $M=\CP^2\times\CP^2\times\CP^2$; the table in (ii) shows the geometry corresponding to the different stabilizers, while in (i) we see the adjacencies of the strata.  The (a) and (b) refer in each case to two different geometry types for the same stabilizer, and hence different components of the corresponding fixed point space. (Note that strata marked with (a) contain the vertex $a$ in their image, the strata marked with (b) contain the vertex $b$ in their image, and the image of the $\TT^2_{(c)}$-strata consists of the points $c_j$.)}
\label{fig:strata}
\end{figure}
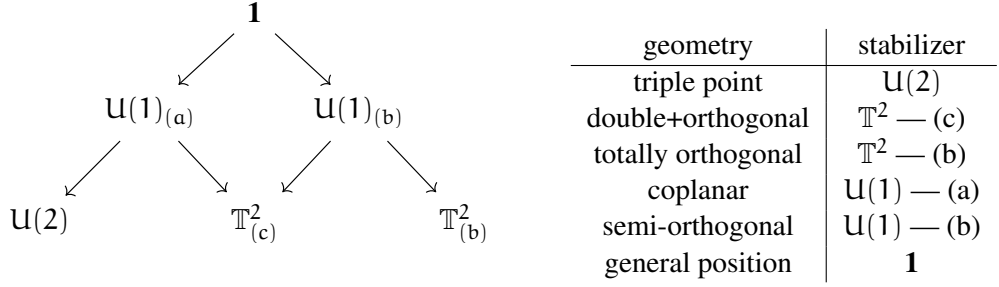

We distinguish a few special configurations in $M$, which have stabilizers given in Figure\,\ref{table:stabilizers}: 

\begin{definition}\label{def:configurations}
A \defn{triple point} in $M$ is where all three elements of $\CP^2$ coincide, a \defn{double point} is where two coincide and the third is different, a \defn{double+orthogonal point} is a double point where the third point is orthogonal to the two coincident ones, a \defn{coplanar point} is where the three points lie in a common copy of $\CP^1$,  a \defn{semi-orthogonal point} is where one point is orthogonal to the other two, which are distinct, and a \defn{totally orthogonal point} is one where all three points of $\CP^2$ are mutually orthogonal. 
\end{definition}

Recall from \cite{MS19a} that the possible stabilizer subgroups for 3 points on $\CP^2$ is as in Figure\,\ref{table:stabilizers}.  In particular, we distinguish the points in $\Delta(M)$ arising from points with stabilizer equal to a (maximal) torus: 
$$a=\J(e_j,e_j,e_j),\quad b=\J(e_i,e_j,e_k),$$
$$c_1=\J(e_i,e_j,e_j),\quad c_2=\J(e_j,e_i,e_j),\quad c_3=\J(e_j,e_j,e_i),$$
where $i,j,k$ are distinct, and are chosen so that the corresponding point lies in our chosen $\tt^*_+$.  Thus, $a$ is the image of a triple point, $b$ of a totally orthogonal point, while the $c_j$ are images of  `double+orthogonal' points.  Moreover, with reference to Figure\,\ref{fig:strata}, the points $a,c_1,c_2,c_3$ are images of points in $\Fix(U(1),M)_{(a)}$ (they are all coplanar configurations), while $b,c_1,c_2,c_3$ are images of points in $\Fix(U(1),M)_{(b)}$.

Before progressing to the statement of the classification of reduced spaces, let us examine this fixed point space of $U(1)$ and its momentum map. This subgroup consists of matrices given in \eqref{eq:subgroups}, and the set of fixed points in $\CP^2$ has two components \eqref{eq:fixed point spaces}.  It follows that $\Fix(U(1),M)$ has $2^3=8$ connected components. In this discussion, we identify $\CP^1$ with the subspace of $\CP^2$ orthogonal to $e_3$.  

The two components of interest of this fixed point space are,
\begin{equation}\label{eq:U(1)FPS}
\Ma = \CP^1\times\CP^1\times\CP^1,\quad\text{and}\quad 
\Mb = \CP^1\times\CP^1\times\{e_3\}.
\end{equation}
(We only consider these components of the fixed point spaces as the others are either permutations of the components of $M_{(b)}$, or are better studied as points in the fixed point space of a conjugate copy of $U(1)$.) 
The symplectic submanifold $\Ma$ consists of coplanar configurations, while $\Mb$ consists of semi-orthogonal ones.  The restriction of the momentum map to each is as follows. With 
$\CP^1=\{[x:y:0]\in\CP^2\}$, 
\begin{equation}\label{eq:Ja}
\Ja(m) =\sum_{j=1}^3\Gamma_j\begin{pmatrix}
|x_j|^2 -\frac13& x_j\overline{y_j}&0\cr
\overline{x_j}\,y_j&|y_j|^2-\frac13&0\cr
0&0& -\frac13
\end{pmatrix},
\end{equation}
while for $\Mb$, the restriction of the momentum map is
\begin{equation}\label{eq:Jb}
\Jb(m) =\sum_{j=1}^2\Gamma_j\begin{pmatrix}
|x_j|^2 -\frac13& x_j\overline{y_j}&0\cr
\overline{x_j}\,y_j&|y_j|^2-\frac13&0\cr
0&0& -\frac13
\end{pmatrix} +
\Gamma_3\begin{pmatrix}
-\tfrac13&0&0\cr 0&-\tfrac13&0\cr0&0&\tfrac23
\end{pmatrix},
\end{equation}

\begin{remark}\label{rmk:SO(3)}
In both cases, $\Ma$ and $\Mb$ are invariant under the subgroup $U(2)$ (equal to the normalizer of $U(1)$ in $SU(3)$), and the momentum maps $\Ja$ and $\Jb$ can be identified with the momentum maps for this $U(2)$ action. Moreover, $U(1)\subset U(2)$ acts trivially, giving an effective action of $U(2)/U(1)\simeq SO(3)$ on $\Ma$ and $\Mb$.  The image of these momentum maps is then contained in an affine copy of $\so(3)^*$ in $\su(3)^*$.  After identifying $\CP^1$ as the Riemann sphere, the two momentum maps can be identified with those for the $SO(3)$ actions on $\Ma\simeq S^2\times S^2\times S^2$ and $\Mb\simeq S^2\times S^2$. For $\Ma$ the momentum map can be rewritten as,
\begin{equation}\label{eq:Ja-SO3}
\Ja(m) = \mu_0 \ + \ 
\sum_{j=1}^3\Gamma_j\begin{pmatrix}
s_j& x_j\overline{y_j}&0\cr
\overline{x_j}\,y_j&-s_j&0\cr
0&&0
\end{pmatrix},
\end{equation}
where $\mu_0=\tfrac16\bigl(\sum_j\Gamma_j\bigr)\diag[1,1,-2]$ and  $s_j=\frac12(|x_j|^2-|y_j|^2)$. 
The first term is a constant while the image of the second lies in $\su(2)^*\simeq\so(3)^*$.
\end{remark}

\begin{figure}
{\begin{subfigure}{0.2\textwidth}
{\includegraphics[height=30mm]{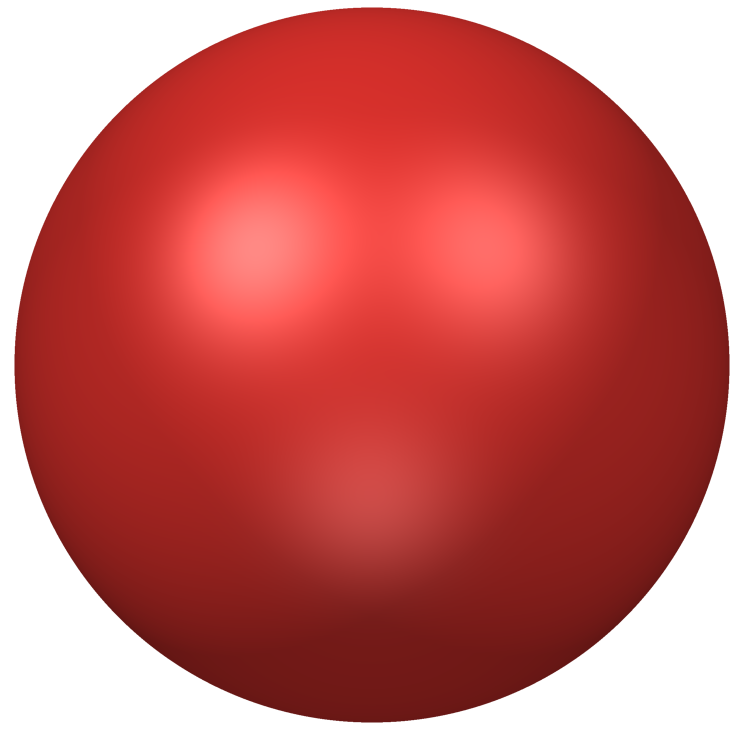}}
\subcaption{Sphere}
\end{subfigure}}\hfill
{\begin{subfigure}{0.2\textwidth}
{\includegraphics[height=30mm]{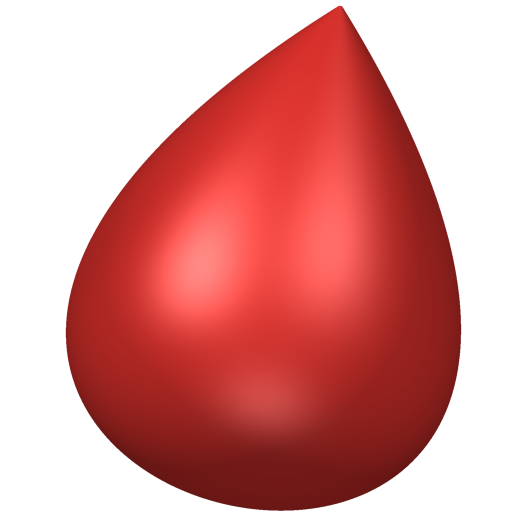}}
\subcaption{pointed sphere}
\end{subfigure}}\hfill
{\begin{subfigure}{0.2\textwidth}
{\includegraphics[height=30mm]{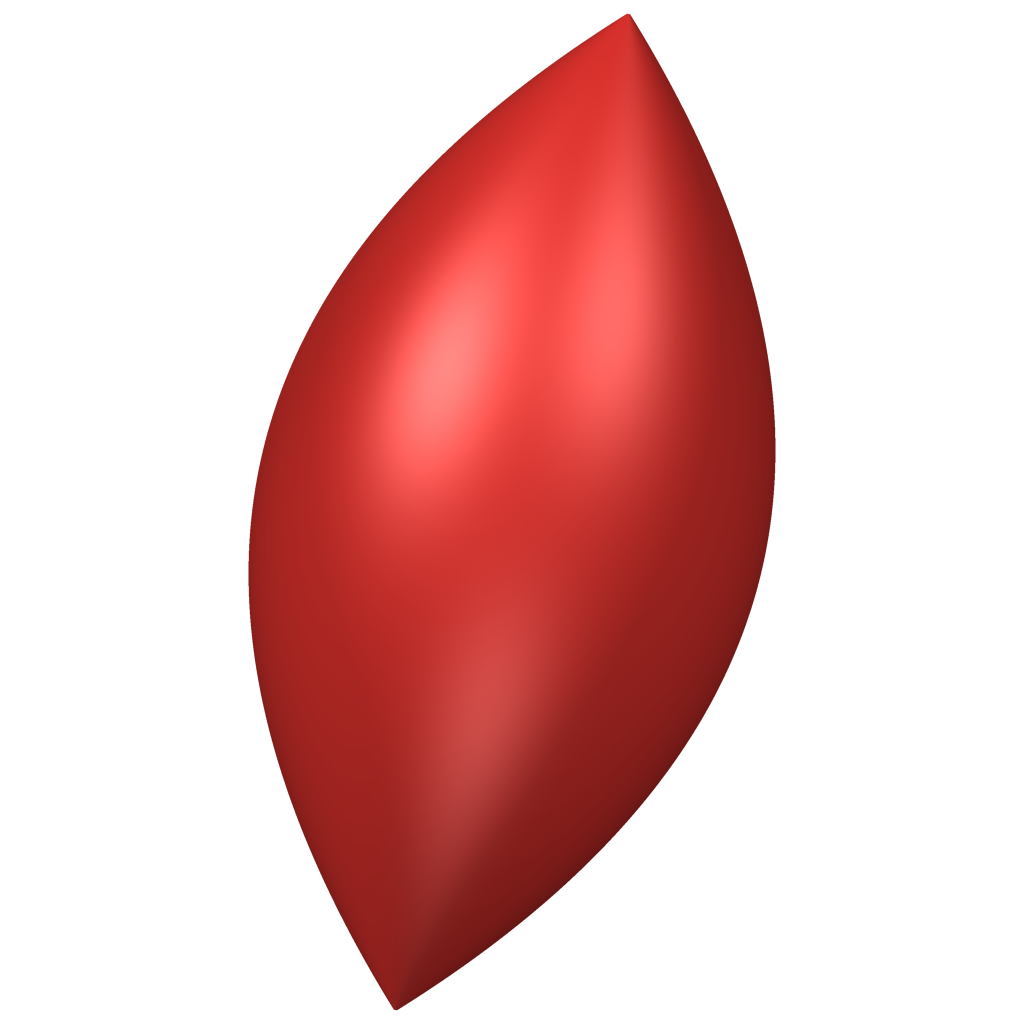}}
\subcaption{\raggedright Twice pointed sphere}
\end{subfigure}}\hfill
{\begin{subfigure}{0.2\textwidth}
{\includegraphics[height=30mm]{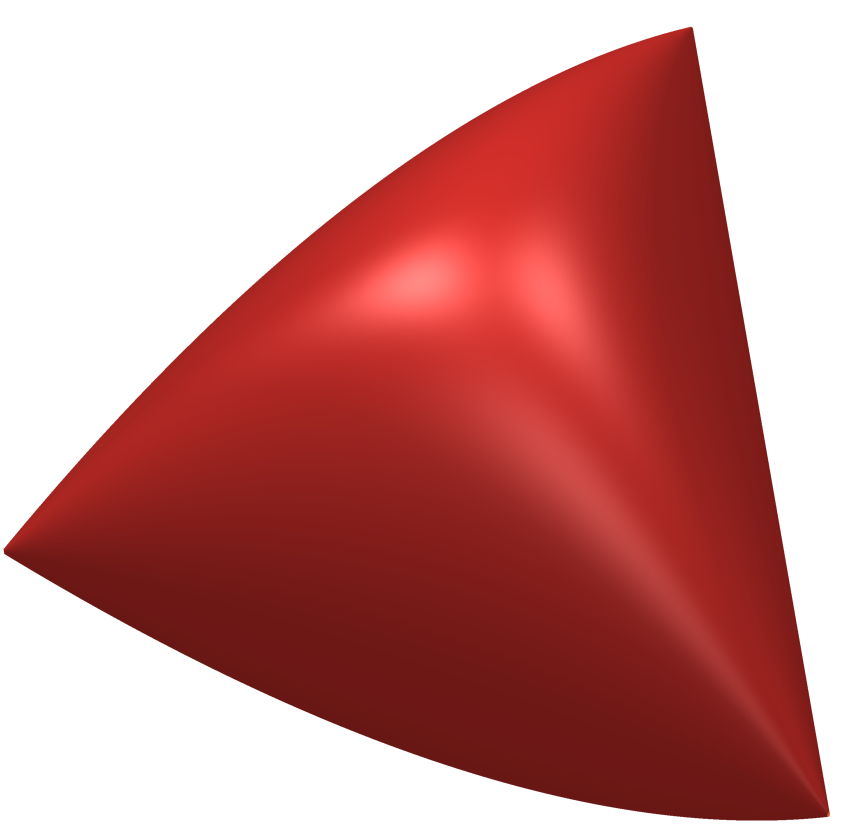}}
\subcaption{\raggedright Thrice pointed sphere}
\end{subfigure}}
\caption{Possible non-trivial reduced spaces for the 3-vortex problem.}
	\label{fig:reduced spaces}
\end{figure}

From the diagrams of the polytopes (eg Fig.\,\ref{fig:polytopes+reduction} below, and those of \cite{MS19a}), one observes that every edge of $\Delta(M)$ lying in $\IntT$ has either $a$ or $b$ as an end-point (and not both).  Accordingly, they are called $a$-edges and $b$-edges respectively. 

\begin{theorem}\label{thm:reduction N=3}
Consider the action of\/ $SU(3)$ on $M=\CP^2\times\CP^2\times\CP^2$, with symplectic form given by \eqref{eq:symplectic form} with non-zero vortex strengths $\Gamma_1,\Gamma_2,\Gamma_3$.   For $\mu\in\Delta(M)$, the reduced space $M_\mu$ is diffeomorphic to a point, a smooth sphere or a `pointed' sphere, as follows,

\begin{center}
\begin{tabular}{llc}
Type of reduced space & conditions on $\mu\in\Delta(M)$ \\
\toprule
$M_\mu$ is a point & vertex not equal to $a$ & (A)\\
	& $a$ when end-point of an edge in a wall  & (A) \\
	& in an edge in a wall  & (A) \\
	& in a $b$-edge  & (A) \\
\midrule
$M_\mu$ is a smooth sphere &  regular value of\/ $J$ in interior of polytope & (B,C) \\
	& point on an $a$-edge distinct from $c_j$ & (C) \\
\midrule
$M_\mu$ is a singular sphere & $\mu$ lies on an `interior edge' & (B,C) \\
	& $\mu=c_j$ lying in an $a$-edge & (C) \\
	& $\mu=a$ when  $\Delta(M)\cap\textrm{Wall}=\{a\}$  & (D) \\
\bottomrule
\end{tabular}
\end{center}

\noindent (The final column refers to which part of the proof refers to which case.)

The reduced space $M_a$ is therefore of dimension 0 if and only if the vortex strengths $\Gamma_j$ are such that the polytope is of Type $A,B,$ or $D$ or one of the associated transitions. See also Figure\,\ref{fig:polytopes+reduction}.

Furthermore, for generic $\Gamma_j$ the singular spheres are spheres with a single singular point as in Figure\,\ref{fig:reduced spaces}(ii). However, in the following cases the reduced space has more than one singular point (but is still homeomorphic to a sphere, see Figs\,\ref{fig:reduced spaces}(iii,iv)):
\begin{itemize}
\item [$\Gamma_1=\Gamma_2\neq\pm\Gamma_3$] (Types AA, HH, GG, GG$_0$, BB, DD, DD$_0$) In these cases $c_1=c_2$ and $M_{c_1}$ has 2 singular points;
\item [$\Gamma_1=\Gamma_2=\Gamma_3$] (Type AAA) In this case $c_1=c_2=c_3$ and $M_{c_1}$ has 3 singular points;
\item [$\Gamma_1=\Gamma_2=-\Gamma_3$] (Type FGH) In this case $c_1=c_2=a$ and $M_a$ has 3 singular points;
\end{itemize}
Similar statements hold by permuting the indices (the polytope type refers to the labels in the Figures in \cite[Sec.\,4]{MS19a}).
\end{theorem}

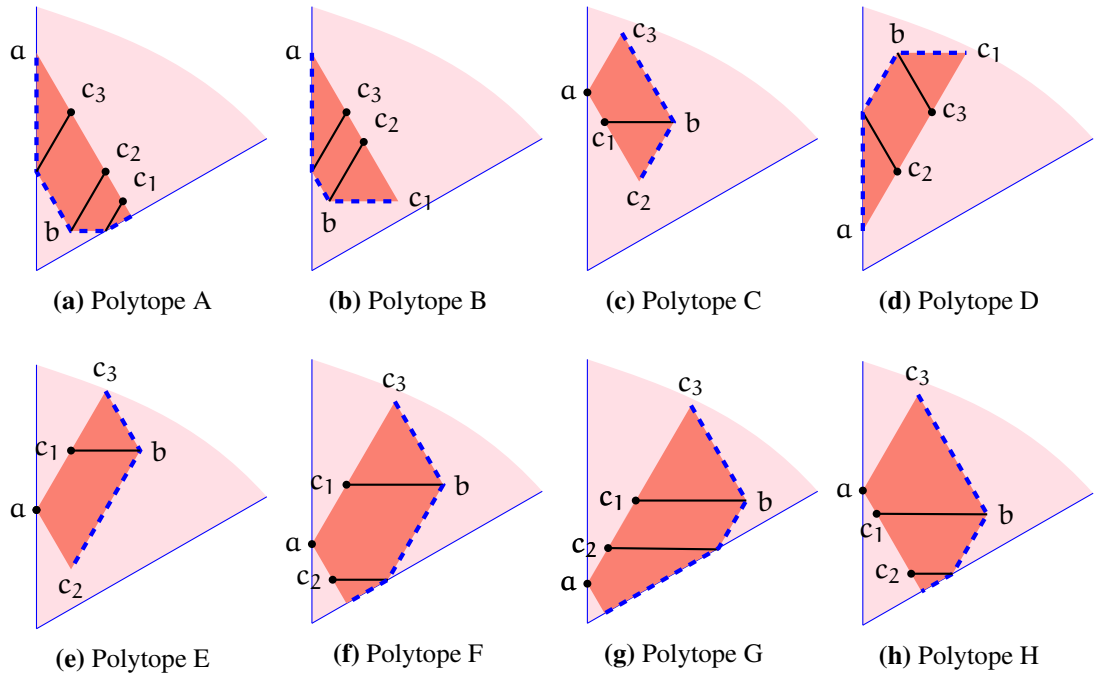
\begin{figure}
	\centering
	
\input{PolytopesA-H+reduction.tex}

\caption{The generic momentum polytopes from \cite{MS19a}, showing the type of reduced space. The salmon coloured regions (including plain boundary points) are where the reduced space is a smooth 2-sphere, the black lines or dots are where it is a once-pointed 2-sphere, and the thick dashed lines represent where the reduced space is a point.}
\label{fig:polytopes+reduction}
\end{figure}

For regular values of the momentum map there are two approaches to proving this. One is by direct calculation, as carried out below, and the other is to use a theorem of Kirwan \cite[Chapter 5]{Kir84a} which implies that since $M$ is compact and simply connected, the regular reduced spaces have vanishing rational cohomology in dimension 1, so when they are of dimension 2 they must be 2-spheres. 

\begin{proof}
The remainder of this section is dedicated to the proof (with the exception of a few remarks at the end).  We proceed case by case; the numbering refers to the final column in the table.  Recall first the `bifucation lemma' for momentum maps, which states that
$$\mathrm{image}(D\J_m) = \gg_m^\circ,$$
where $\gg_m$ is the Lie algebra of the stabilizer of the point $m$, and $\gg_m^\circ$ its annihilator in $\gg^*$. This follows readily from the definition of a momentum map. Recall also that  the fibres of any proper momentum map defined on a connected compact symplectic manifold are connected \cite{Sjamaar}; we use this without further mention below. 

Recall the local normal form at a point $m$ for the momentum map, described in \cite[Sec.~2.1]{MS19a}, 
 $$\J([g,\sigma,v]) = g\left(\mu+\sigma+\J_{N_1}(v)\right)g^{-1},$$
where 
$$[g,\sigma,v]\in Y(\mu,G_m,N_1) \ = \ G \times_{G_m} (\nn\oplus N_1).$$
Here $N_1$ is the symplectic slice at $m$, and $\nn$ can be identified with $\gg_\mu^*\cap\gg_m^\circ$. In the following we freely use results of calculations from \cite{MS19a} for the form of $\J_{N_1}$. 

Recall that a $b$-edge is any edge of the polytope lying in $\IntT$ with $b$ as one end-point.  By the bifurcation lemma, any $m$ for which $J(m)$ belongs to a $b$-edge necessarily has stabilizer containing $U(1)$, for if the action were locally free at $m$ then $\Delta(M)$ would contain a full neighbourhood of $\mu$ in $\tt^*_+$.  Thus $m$ belongs to $\Ma$ or $\Mb$.  Since $(e_1,e_1,e_1)\in\Ma$ it follows that $J(\Ma)$ is an $a$-edge, while $J(\Mb)$ is a $b$-edge (possibly the `interior' $b$-edge shown for example as the horizontal line in Figure\,\ref{subfig:N=3 E}).  Recall from Remark\,\ref{rmk:SO(3)}, that the momentum map $\Jb:\Mb\to\left(\mu_0+\so(3)^*\right)\subset\su(3)^*$ can be identified with the one for the $SO(3)$ action on $S^2\times S^2$, and similarly, $\Ja:\Ma\to\left(\mu_0+\so(3)^*\right)\subset\su(3)^*$ can be identified with the one for the $SO(3)$ action on $S^2\times S^2\times S^2$. 

The proof now consists of a case-by-case analysis, as marked in the right-hand column of the table.

\noindent(A) 
We first deal with some easy cases where a dimension count shows the reduced space is a point.  If $\mu=J(m)$ is a vertex of $\Delta(M)$ in the interior of $\tt^*_+$, then it follows from the bifurcation lemma above that the stabilizer of $m$ contains the maximal torus $\TT$. Now the set of points fixed by $\TT$ is finite, and hence $J^{-1}(\mu)=\{m\}$ and so $M_\mu$ is a single point. This covers $M_b$ and $M_{c_j}$ when $c_j$ is a vertex.  

Now let $\mu\in\IntT$ belong to a $b$-edge of $\Delta(M)$. Every point in the fibre has stabilizer at least $U(1)$, and is contained in $\Mb$ (as described above).  Now $\Mb\simeq S^2\times S^2$, so the generic fibre is of dimension 1, and the reduced spaces are points.

Now suppose $\mu$ is a regular value of the momentum map belonging to a wall of the Weyl chamber.  Then $\dim J^{-1}(\mu) = 12-8=4$, and the group $G_\mu\simeq U(2)$ acts freely on this fibre (freely because $\mu$ is a regular value), and hence $M_\mu$ is a single point. 

Finally, suppose $\mu$ is a vertex contained in a wall, but distinct from $a$. Such a point is the intersection of an edge of $\Delta(M)$ with a wall of the Weyl chamber.  There are two cases.  Firstly, suppose the edge in question is a $b$-edge. Then  the reduced space is a point as for any point on a $b$-edge.  If on the other hand, the edge is an $a$-edge (as for polytope H for example), the preimage $\J^{-1}(\mu)$ is contained in $\Ma\simeq \left(S^2\right)^3$.  This preimage is of dimension $6-3=3$, with an effective action of $G_\mu=SU(2)\subset U(2)$, and hence the reduced space is a point.

\smallskip\noindent(B) Consider a neighbourhood of the vertex $b$, assuming the three vortex strengths to be distinct (otherwise $b\not\in\IntT$).  Let us assume $\Gamma_1 > \Gamma_2 > \Gamma_3$, in which case $\J(m)\in\tt^*_+$ for $m=(e_1,e_2,e_3)$. 
Using  \cite[Eq.\,(4.2)]{MS19a}, we have $N_1=\C^3$ and the momentum map for the action of the maximal torus is 
 $$ \J_{N_1}(w,u,v) = 
	\frac{\Gamma_3}{\Gamma_2}(\Gamma_2-\Gamma_3)|v|^2\alpha_1 \ + \  
	\frac{\Gamma_1}{\Gamma_3}(\Gamma_3-\Gamma_1)|w|^2\alpha_2 \ + \  
	\frac{\Gamma_2}{\Gamma_1}(\Gamma_1-\Gamma_2)|u|^2\alpha_3,
$$
where $\alpha_j$ are the roots of $SU(3)$ as shown in Figure\,\ref{fig:the roots}.  (We have replaced $u_2$ by $u$ etc.) Of the coefficients, two are positive and one is negative.  Suppose it is the coefficient of $\alpha_1$ that is negative (the other possibilities are similar). Substituting $\alpha_1=-(\alpha_2+\alpha_3)$  gives an expression of the form 
\begin{equation}\label{eq:b localMM}
\J_{N_1}(w,u,v) = (A|v|^2+B|w|^2)\alpha_2+(A|v|^2+C|u|^2)\alpha_3,
\end{equation}
with $A,B,C>0$.

\begin{lemma}\label{lemma:b-vertex}
Consider the $\TT$-invariant map $J=J_{N_1}$ defined in \eqref{eq:b localMM} above.  The quotients by $\TT$ of the fibres are as follows:
\begin{enumerate}
\item  $J^{-1}(0,0)=\{(0,0)\}$
\item For $t>0$,  $J^{-1}(t\alpha_2)\simeq J^{-1}(t\alpha_3)\simeq S^1$, and hence the orbit space is a single point.
\item For $s,t>0,\,s\neq t$: the quotient $J^{-1}(s\alpha_2+t\alpha_3)/\TT$ is diffeomorphic to a smooth 2-sphere. 
\item{}For $t>0$, the fibre $J^{-1}(t(\alpha_2+\alpha_3))$ is singular of dimension 4. The orbit space is a once-pointed sphere.  
\end{enumerate}
\end{lemma}

\begin{proof}
(1) and (2) are clear. 

(3) Write $X=J^{-1}(s\alpha_2+t\alpha_3)$.  Since this is a regular value of $J$, this is a smooth manifold of dimension 4.  Moreover, $\TT$ acts freely on $X$ and hence $X$ is a smooth compact surface.  Consider the $\TT$-invariant smooth function $f(u,v,w)=|u|^2$.  This function has two critical $\TT$-orbits: a maximum along $\{(u,v,w)\in X\mid |u|^2=y,\,|v|^2=x-y,\,w=0\}$, and a minimum along $\{(u,v,w)\in X\mid u=0,\, |v|^2=x,\,|w|^2=y\}$. It follows that $f$ descends to a smooth function on $X/\TT$ with two isolated critical points, and hence by a famous theorem of Reeb (see \cite{Milnor}) that $X/\TT$ is a 2-sphere. (Note that Reeb's theorem states that the surface is homeomorphic to the sphere, but in dimension 2 any smooth surface homeomorphic to the sphere is diffeomorphic to it).  

(4) Now apply the previous argument with $s=t>0$.  The function $|u|^2$ is still invariant and has a unique orbit of maxima and minima.  However, the fibre $F^{-1}(t,t)$ is a real algebraic variety of dimension 4 with singular locus equal to $\{(w,u,v)=(0,0,v)\mid A|v|^2=t\}$. This singular orbit coincides with the minimum of $|u|^2$ on the fibre; the mximum occurs where $v=0$, which is a single $\TT$-orbit.  Outside of this singular circle, the action of $\TT$ is free. It follows that the orbit space is a 2-dimensional space which is smooth outside of a single singular point, and that singular point is a cone (that is, diffeomorphic to $x^2+y^2=z^2$ with $z\geq0$ in $\R^3$).  Milnor's proof of Reeb's theorem continues to hold in this setting, showing that the reduced space is the union of two 2-cells (one diffeomorphic to a cone), forming a 2-sphere with a single conical singular point. 
\end{proof}

From this lemma, we deduce: (1) the reduced space $M_b$ is a point, (2) if $\mu$ lies on a $b$-edge then $M_\mu$ is a point   (both of these are already proved above), (3) for regular points of $\Delta(M)$ in regions containing $b$ in their closure the reduced space is a 2-sphere, and (4) over the internal edge emanating from (b) the reduced space is a pointed sphere.

\begin{figure} 
\centering
  { \begin{tikzpicture}[scale=0.2]
\path [fill=pink] (0,0) -- (0,10) .. controls (3,9) and  (6,8) .. (8.66,5) -- (0,0);
\draw[purple](5,3)
	 node[anchor=south]{ \Large $\mathfrak{t}^*_+$};
\draw [blue,thick,->] (0,0) -- (6,0) node[anchor=north] {$\alpha_1$};
\draw [rotate=-120,blue,thick,,->] (0,0)  -- (6,0) node[anchor=north] {$\alpha_2$};
\draw [rotate=120,blue,thick,->] (0,0)  -- (6,0) node[anchor=south] {$\alpha_3$};
 \end{tikzpicture}}
\caption{The roots $\alpha_1,\alpha_2,\alpha_3$ and positive Weyl chamber of $SU(3)$, see \cite{MS19a}}
\label{fig:the roots}
\end{figure}
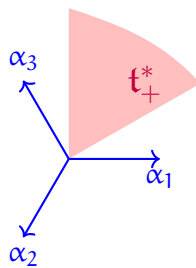

\smallskip\noindent(C) Consider a neighbourhood of $\mu=c_j$ when $c_j$ does not lie in a wall of the Weyl chamber.  Refer to the expressions for the symplectic slice momentum map $\J_{N_1}$ for $c_1,c_2$ and $c_3$ in \cite[Eqs\,(4.3--4.5)]{MS19a}.  For example, for $c_1$ it is shown that 
$J_{N_1}(u,v,w) = R\alpha_1+S\alpha_3$
where $R= - \frac{\Gamma_2}{\Gamma_3}(\Gamma_2+\Gamma_3)|w|^2$, and 
$$S =  \left(\frac{\Gamma_1}{\Gamma_2}(\Gamma_1-\Gamma_2)|v|^2+
	 \frac{\Gamma_1\Gamma_3}{\Gamma_2}(uv+\overline{uv})+
	 \frac{\Gamma_3}{\Gamma_2}(\Gamma_3+\Gamma_2)|u|^2\right),
$$
where we have replaced $u_3$ by $u$ etc.. The expressions for $c_2,c_3$ are similar.  If, as we assume, $c_1$ is not in the wall then $S$ is non-degenerate and $R$ is non-zero. Let us suppose for sake of argument that $R\geq0$ (i.e., $\Gamma_2\Gamma_3(\Gamma_2+\Gamma_3)<0$).  There are thus two cases to consider: the real quadratic form $S$ is definite or indefinite.  

Consider first the definite case. In this case $J_{N_1}^{-1}(0)=0$, and the reduced space is a point (as we already know from part (A): in this case $c_1$ is a vertex of $\Delta(M)$).  Similarly, for $t>0$,  $J_{N_1}^{-1}(t\alpha_1)$ is a circle (with $u=v=0$), and the reduced space is again a point (as we also know already: these points lie on a $b$-edge). On the other hand, $J_{N_1}^{-1}(t\alpha_3)$ is diffeomorphic to a 3-sphere (it's an ellipsoid of dimension 3 in $\C^2$) on which there is an action of $U(1)$ and the quotient is a 2-sphere.  Note from the figures in \cite{MS19a} that this edge joins $c_1$ to $a$, so is an $a$-edge (though it passes through at least one of the other $c_j$ before reaching $a$). 
Now consider $J_{N_1}^{-1}(s\alpha_1+t\alpha_3)$ with $s,t>0$. This is diffeomorphic to $S^1\times S^3$, with a free action of $\TT^2$, and again the quotient is $S^2$. 

Now suppose that $S$ is indefinite, in which case it is of signature zero (2 positive eigenvalues and 2 negative). In this case $c_1$ lies on an $a$-edge of $\Delta(M)$ (here parallel to $\alpha_3$).  The inverse image of this edge is $\Ma$, so first we restrict to this.  For $t\neq0$, the subset $S^{-1}(t)$ is a smooth 3-dimensional submanifold of $\C^2$ (a hyperboloid) with a free action of $U(1)$ whose quotient is therefore a smooth non-compact 2-dimensional surface. This surface is an open subset of the 2-sphere described above (for an $a$-edge), when $c_1$ (or other $c_j$) is a vertex.  On the other hand, the set of solutions to $S=0$ is a 3-dimensional conical subspace of $\C^2$ (in fact a cone over a 2-torus since $S$ has index 2). After factoring out by the remaining $U(1)$-action (here $U(1)=SO(3)_\mu$ for non-zero $\mu\in\so(3)^*$), one concludes that the reduced space is 2-dimensional, with a single conical point.   It follows from the argument in the lemma below that the reduced space is a topological sphere with a single singular point at $m$.

\smallskip\noindent(D) There remains to consider $M_a$.  Since (for all $\Gamma_j$) the polytope $\Delta(M)$ does not contain a full neighbourhood of $a$, every point in $\J^{-1}(a)$ has non-trivial stabilizer.  Up to conjugacy, the stabilizer subgroups and corresponding strata are listed in Figure\,\ref{fig:strata}. 

 It therefore suffices to consider $\Ja:\Ma\to\su(2)^*$ (see also Remark\,\ref{rmk:SO(3)}).  Now, in  a neighbourhood of the triple point $m=(e_1,e_1,e_1)$, the full momentum map is determined by the action on the symplectic slice (see \cite[Eq.~(4.6)]{MS19a}) 
$$\J_{N_1} = 
\begin{pmatrix}
-\sum_j\Gamma_j(|v_j|^2+|w_j|^2) & 0 & 0\\
0 & \sum_j\Gamma_j|v_j|^2 & \sum_j\Gamma_j \overline{v_j}w_j \\
0 & \sum_j\Gamma_jv_j\overline{w_j} &\sum_j\Gamma_j|w_j|^2 
\end{pmatrix},
$$
subject to $\sum_j\Gamma_jv_j=\sum_j\Gamma_jw_j=0$.  Restricting to $\Ma$ imposes $w_j=0$, and hence, locally in $\Ma$,
$$\J_{N_1,(a)}(v_1,v_2,v_3) = \bigl(\sum_j\Gamma_j|v_j|^2\bigr)\diag\bigl[-1,\;1,\;0\bigr]
$$
again, subject to $\sum_j\Gamma_jv_j=0$. This quadratic form on $\C^2$ (after putting $\sum_j\Gamma_jv_j=0$) is definite if and only if $\Gamma_1\Gamma_2\Gamma_3(\Gamma_1+\Gamma_2+\Gamma_3)>0$  (see \cite[Lemma 4.3]{MS19a}), in which case $M_a$ is just a point. 

 If instead this expression is negative, then the zero-set of $\J_{N_1,(a)}$ is a cone over a 2-torus, and the quotient by $U(1)$ gives a cone over a circle, which is an ordinary conical point in a 2-dimensional surface. The argument of the lemma below shows than that  $M_a$ is a topological 2-sphere, with singular point at $m$.

\begin{remark}\label{rmk:J_Z}
In the study of the reduced space $M_a$, it is instructive to consider the restriction of the action of $U(2)$ on $N_1$ to the action of the centre $Z\simeq U(1)$ of $U(2)$, corresponding to the wall containing the point $a$.  The momentum map for the $Z$-action is the composite of $J_{N_1}$ with the trace of the $U(2)$ part, that is
$$\J_Z(v,w) = \sum_j \Gamma_j\left(|v_j|^2+|w_j|^2\right),$$
on the symplectic slice $\sum\Gamma_jv_j=\sum\Gamma_jw_j=0$. 
As above, this is definite if and only if $\Gamma_1\Gamma_2\Gamma_3(\Gamma_1+\Gamma_2+\Gamma_3)>0$.  In this case, $J_Z^{-1}(0)=0$ so the reduced space is a point (already for this action of $Z$, without needing to consider the full action of $U(2)$). Compare this with the projection of the polytope to the wall containing $a$, one sees that indeed this is definite for the polytopes of type A, B and D. 
\end{remark}

We end the proof with a lemma containing the argument showing that indeed the singular reduced spaces are topological spheres with up to 3 conical singular points. 

\begin{lemma}\label{lemma:singular spheres}
Suppose $\mu\in\Delta(M)$ is such that $M_\mu$ is of dimension 2 but has a singular point. Then it is homeomorphic to a 2-sphere with the number of singular points given in the theorem. 
\end{lemma}

\begin{proof} 
For such $\mu$, the singular points occur at points $m$ for which $G_m$ is strictly larger than nearby points in $J^{-1}(\mu)$.   Suppose for simplicity there is just one $G_\mu$-orbit of singular points $m$ in $J^{-1}(\mu)$.   (If there is more than one `singular' $G_\mu$ orbit of such points, then this argument should be repeated for each.)    
Let $U$ be a (small) G-invariant neighbourhood of $m$ in $M$.  Then by $G$-openness of $J$ (see \cite{Sjamaar, MT03}) $J(U)$ is a neighbourhood of $\mu$ in $\Delta(M)$.  Let $L$ be a compact line segment in $J(U)\cap\tt^*_+$ with one end-point at $\mu$ and otherwise contained in the set of $\mu$ for which $M_\mu$ is a smooth sphere.  

Consider the restriction of the orbit momentum map, $\j_L:J^{-1}(L)/\TT\to L$, where we identify $J^{-1}(L)/\TT$ with $J^{-1}(G\cdot L)/G$.  

Within $U$, the local calculations show the fibre of this map is an open disc, except over $\mu$ where the fibre is a cone, which is a topological disc. 
Outside of $U$ this map is a submersion, so is a smooth fibration with fibre a closed disc (the complement in $S^2$ of the open disc arising from the interection of the fibre with $U$).  Gluing these together at $\mu$ shows that the full fibre over $\mu$ is a topological sphere with a single conical singularity at $m$. 
\end{proof}

\paragraph{Transition polytopes} Some of the arguments above extend to the cases where $\Gamma_j$ are `transition' cases, that is, in the boundaries between the regions A, B, \dots, H.  Here we discuss some of these.

\medskip

\noindent\underline{$\Gamma_1=\Gamma_2\neq\Gamma_3$}:  the vertex $b$ is contained in a wall of the Weyl chamber, and $c_1=c_2$. In that case, $b$ lies in a wall of the Weyl chamber.  The Witt-Artin decomposition at $m=(e_1,e_2,e_3)$ satisfies $\dim T_0=2=\dim N_0$ and hence the symplectic slice is only of dimension 4 instead of 6. The momentum map on the symplectic slice (which we identify with $\C^2$) is
$$  \J_{N_1}(w,v) = 
	\frac{\Gamma_3}{\Gamma_1}(\Gamma_1-\Gamma_3)|v|^2\alpha_1 \ + \  
	\frac{\Gamma_1}{\Gamma_3}(\Gamma_3-\Gamma_1)|w|^2\alpha_2. $$
(cf. \cite[Eq,\,(4.2)]{MS19a}, with $\Gamma_1=\Gamma_2$).  It is clear that $\J_{N_1}^{-1}(0)$ is just the origin, and this (or its quotient by $G_m$) provides a local model for the reduced space over $\mu=b$, which is therefore just a point.  The reduced space over $c_1=c_2$ will be again of dimension 2, but with two singular points:  it is a twice-pointed sphere. Other reduced spaces will be as usual. 

\medskip

\noindent\underline{$\Gamma_1=\Gamma_2=\Gamma_3$}:  
In this case the polytope is of Type AAA  (see \cite[Fig.\,4.8b]{MS19a}),  $b=0$ and $c_1=c_2=c_3$.  For $M_b$,  the Witt-Artin decomposition at $m=(e_1,e_2,e_3)$ has $\dim T_0=\dim N_0=6$, and hence $N_1=0$.  Then the MGS normal form is
$$\J_Y([g,\sigma]) = g\sigma g^{-1}$$
with $\sigma\in N_0 \simeq \tt^\circ\subset\su(3)^*$.  It follows that $\J^{-1}(0)$ is just a single orbit, and again the reduced space $M_b$ is a single point. On the other hand, by the lemma above and the calculations in part (C),  the reduced space at $c_1=c_2=c_3$ is a thrice-pointed sphere (Fig.\,\ref{fig:reduced spaces}(d)).  The other reduced spaces will be as usual. 

\medskip

\noindent\underline{$\Gamma_1=\Gamma_2+\Gamma_3$}:  In this case $c_1$ lies in a wall of the Weyl chamber.  The reduced space $M_{c_1}$ remains a single point; other reduced spaces are as usual. 

\medskip

\noindent\underline{$\Gamma_1+\Gamma_2+\Gamma_3=0$}:  In this case $a=0$, and at $(e_1,e_1,e_1)$ one finds $\dim N_1=4$. If $\Gamma_2\neq\Gamma_3$ then $N_1$ can be parametrized by $v_1,w_1$ in which case
$$J_{N_1}(v_1,w_1) = \frac{\Gamma_1\Gamma_2(\Gamma_1+\Gamma_2)}{(\Gamma_2-\Gamma_3)^2}
	\begin{pmatrix}
	-|v_1|^2-|w_1|^2&0&0 \cr 0 & |v_1|^2 & v_1\overline{w_1} \cr 0 & \overline{v_1}w_1 & |w_1|^2
	\end{pmatrix}.
$$
In this case, with our assumptions it is not possible for $\Gamma_1+\Gamma_2=0$ (see \cite[Figure~4.1]{MS19a}), and hence $J_{N_1}^{-1}(0)=0$, and consequently in this case $M_a$ is just a point. (If instead $\Gamma_2=\Gamma_3$ one can similarly parametrize $N_1$ by $v_2,w_2$ and arrive at similar conclusions). 
\medskip

\noindent\underline{$\Gamma_2+\Gamma_3=0$}:  Here $a=c_1$ and the reduced space $M_a$ is now a twice-pointed sphere. The other reduced spaces will be as usual. 

\medskip

Other transition cases can be treated similarly.  Herewith endeth the proof of the theorem. 
\end{proof}

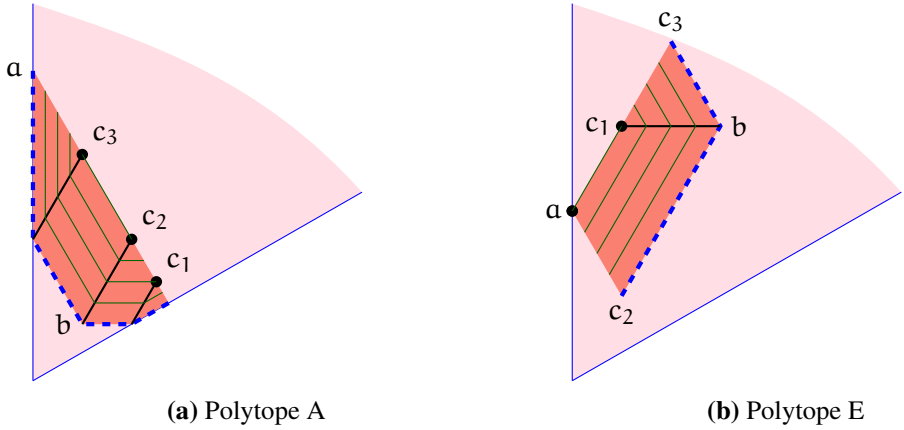
\begin{figure}
\centering 

{\begin{subfigure}{0.45\textwidth}
\begin{tikzpicture}[scale=0.5] 
\WeylChamber
	\fill[fill=Salmon] (0, 8.25) -- (3.572, 2.062) -- 
	 (2.598, 1.5) -- (1.3, 1.5) -- (0, 3.75) -- (0, 8.25);
	\draw[ultra thick,dashed,blue] (3.572, 2.062) -- (2.598, 1.5) -- 	(1.3, 1.5) -- (0, 3.75) -- (0, 8.25);
	\draw[thick] (1.299, 6) -- (0, 3.75);
	\draw[thick] (2.598, 3.75) -- (1.3,1.5);
	\draw[thick] (3.248, 2.625) -- (2.598, 1.5);
	 \draw (0, 8.25) node[anchor=east]{$a$};
	 \draw (1.3, 1.5) node[anchor=east]{$b$};
	 \fill (1.299, 6) circle [fill,radius=0.15] node[anchor=south west]{$c_3$};
	 \fill (2.598, 3.75) circle [fill,radius=0.15] node[anchor=south west]{$c_2$};
	 \fill (3.248, 2.625) circle [fill,radius=0.15] node[anchor=south west]{$c_1$};
	\draw [color=DarkGreen] (.32475, 7.6875) -- (.32475, 4.3125) -- (1.62450, 2.0625) -- (2.9230, 2.0625) -- (3.4100, 2.3435) ; 
	\draw [color=DarkGreen]  (.6495, 7.125) -- (.6495, 4.875) -- (1.9490, 2.625) -- (3.248, 2.625) ;
	\draw [color=DarkGreen]  (.97425, 6.5625) -- (.97425, 5.4375) -- (2.27350, 3.1875) -- (2.923, 3.1875);
	\draw [color=DarkGreen]  (1.299, 6) -- (2.598, 3.75) ;
 \end{tikzpicture}
\caption{Polytope A}
\label{subfig:Chern N=3 A}
\end{subfigure}}
\quad
{\begin{subfigure}{0.45\textwidth}
\begin{tikzpicture}[scale=0.5] 
\WeylChamber
	\fill[fill=Salmon] (0,4.5) -- 
	(2.598, 9) node [anchor=south]{$c_3$} --
	(3.897, 6.7505) node [anchor=west]{$b$} --
	(1.299, 2.25) node [anchor=north]{$c_2$} -- (0,4.5);
	\draw[ultra thick,dashed,blue] (2.598, 9)  -- 	(3.897, 6.7505)  --(1.299, 2.25);
	\draw[thick] (1.299, 6.75) -- (3.897, 6.7505);
	\fill (0,4.5)  circle [radius=0.15] node[anchor=east]{$a$};	
	\fill (1.299, 6.75) circle [radius=0.15] node[anchor=east]{$c_1$};
	\fill (0,4.5) circle [radius=0.15] ;
	\draw [color=DarkGreen]   (2.27325, 8.4375) -- (3.24750, 6.750375) -- (.97425, 2.8125) ;
	\draw [color=DarkGreen]   (1.9485, 7.875) -- (2.5980, 6.75025) -- (.6495, 3.375) ;
	\draw [color=DarkGreen]   (1.62375, 7.3125) -- (1.94850, 6.750125) -- (.32475, 3.9375) ;
	\draw [color=DarkGreen]   (0,4.5) -- (1.299, 6.75) ;
 \end{tikzpicture}
\caption{Polytope E}
\label{subfig:Chern N=3 E}
\end{subfigure}}
	
\medskip

\caption{The green lines are contours of constant volume of reduced space; see Remark\,\ref{rmk:Chern} and Fig.\,\ref{fig:polytopes+reduction} for the key.}
\label{fig:Chern}
\end{figure}

\begin{remark}\label{rmk:Chern}
The Duistermaat-Heckman theorem \cite{DH82} (valid also for Hamiltonian actions of non-Abelian groups) states that the cohomology class of the symplectic form on the reduced space $M_\mu$ depends linearly on $\mu$ in each connected component of the set of regular values in $\Delta(M)$: explicitly, if $\mu_1$ and $\mu_2$ belong to the same connected component of the set of regular values, then 
$$[\omega_{\mu_1}]-[\omega_{\mu_2}]=\left<c,\,\mu_1-\mu_2\right>,$$
where $c$ is the Chern class of the bundle $J^{-1}(\mu)\to M_\mu$, interpreted as an element of $\tt$ (as described in \cite{DH82}).  In this example, since every regular component of $\Delta(M)$ contains an edge where the reduced space is a point, the Chern class must be orthogonal to (or annihilate) that edge.  And it is non-zero since in the interior the reduced space has non-zero symplectic volume. In Figure\,\ref{fig:Chern} we show two examples; the thin green lines are contours of constant volume of the reduced space; on each region, they are parallel to the one edge where the reduced space is a point. Note also that in Polytope A in Fig.\,\ref{fig:Chern}(a), over the line segment with end-points $c_2$ and $c_3$, the fibration has fibre $S^2$ and with constant volume; the Chern class must therefore vanish for that fibration, which is therefore the trivial fibration $S^2\times U(1)\to S^2$. The same is true over the segment from $a$ to $c_1$ in Polytope E in the figure, and over similar segments in the other polytopes.  See \cite{GS-birational} for an analysis of the relation between the reduced symplectic forms on adjacent regions in terms of blow-ups and blow-downs. 
\end{remark}

\section{Dynamics and relative equilibria}
\label{sec:dynamics}
We now consider aspects of the dynamics for the generalized point vortex system, described in the introduction, with 2 or 3 point vortices, and in particular possible relative equilibria.  We assume the pairwise interaction is governed by an $SU(3)$ invariant Hamiltonian, 
$$h_0:\CP^2\times\CP^2\setminus\Delta \longrightarrow \R,$$
as described in the introduction (here $\Delta$ is the diagonal).  One may also consider interactions allowing collisions, where $h_0$ extends to a smooth function on $\CP^2\times\CP^2$.  We write $M^\circ$ for the open subset of  $M=\CP^2\times\dots\times\CP^2$ obtained by removing the large diagonal (or collision set). Such an invariant function on $M$ or $M^\circ$ will be a smooth function of the distance defined in \eqref{eq:metric}.

Given the symplectic form $\Omega=\sum_j \Gamma_j\pi_j^*\omega_0$, as described in the introduction, the dynamics is given by Hamilton's equation $\dot x = X_H(x)$
where $X_H$ is the vector field satisfying $d H = \Omega(-,X_H)$, and $H:M\to\R$ is given by
\begin{equation}\label{eq:pairwise interaction}
h(x_1,\dots,x_N) = \sum_{i<j} \Gamma_i\,\Gamma_j\,h_0(x_i,x_j).
\end{equation}

\subsection{Relative equilibria and allowed velocity vectors}
We consider for the moment the general setting of a $G$-invariant Hamiltonian system on a symplectic manifold $\cP$.  See for example \cite{MarsRati94} or \cite{JM-peyresq98} for definitions.  For $\xi\in\gg$, the associated vector field on $\cP$ is denoted by $\xi_\cP$ and, given a Hamiltonian function $H:P\rightarrow\R$, the associated vector field is denoted by $X_H$. 

A relative equilibrium is a trajectory that lies in a group orbit or, what is essentially the same, a  group orbit which is invariant under the dynamics. The fact that the trajectory lies in the group orbit means that throughout this orbit, the Hamiltonian vector field is tangent to this orbit, so that $x$ lies on a relative equilibrium if and only if there is a $\xi\in\gg$ for which $X_H(x)=\xi_\cP(x)$.  Such a value of $\xi$ is an \defn{angular velocity} of the relative equilibrium in question. Using the symplectic form this becomes $dH_x=\xi\cdot D\J_x$ and so is equivalent to requiring $x$ to be a critical point of $H_\xi=H-\xi\cdot J$.  If the level set $J^{-1}(\mu)$ is non-singular, then it follows that $x\in J^{-1}(\mu)$ lies on a relative equilibrium if and only if $x$ is a critical point of the restriction of $H$ to $J^{-1}(\mu)$.  Thus the relative equilibria are given by constrained critical points of $H$ in much the same way that equilibria are given by ordinary critical points. 

If the point $x$ has a particular symmetry, then so must the angular velocity $\xi$, as the following result shows. 

\begin{proposition}
Let $x\in \cP$ be a relative equilibrium for a $G$-invariant Hamiltonian system $H$. Then 
\begin{equation}
X_H(x) \in R_0 := (\gg_\mu\cdot x)^{G_x}\subset T_xM. 
\end{equation}
\end{proposition}

We call this subspace $R_0$ the space of \defn{allowed velocity vectors}.  We emphasize this is only a restriction on the velocity if we know that $x$ is a relative equilibrium. Since the kernel of the map $\gg\to T_x\cP$ given by $\xi\mapsto \xi_\cP(x)$ is precisely $\gg_x$, it follows that
\begin{equation}
R_0\simeq (\gg_\mu/\gg_x)^{G_x}.
\end{equation}

Note that if $R_0=0$ then a relative equilibrium is necessarily a (group orbit of) equilibria.

\begin{proof}
This is a combination of conservation of symmetry (which holds for any symmetric dynamical system) with conservation of momentum. Since $x$ is a relative equilibrium, there is a $\xi\in\gg$ for which $X_H(x)=\xi_\cP(x)$.  By conservation of momentum, $D\J_x(X_H(x))=0$ and hence 
$$\xi_\cP(x) \in \gg\cdot x\cap \ker D\J_x = \gg_\mu\cdot x.$$
Now, for any symmetric dynamical system $\dot x = f(x)$, the vector field is tangent to the fixed point spaces: $f(x)\in T_x(\Fix(G_x,\cP))$.  This latter subspace of $T_x\cP$ is equal to $\Fix(G_x,T_x\cP)$.  Combining these shows that indeed for a relative equilibrium,
$$X_H(x)\in\gg_\mu\cdot x \cap \Fix(G_x,T_x\cP) = (\gg_\mu\cdot x)^{G_x}$$
as reqired.
\end{proof}

\subsection{Dynamics for 2 generalized point vortices}
Since the reduced spaces in this instance are single points (Theorem\,\ref{thm:reduction 2}), every trajectory is a relative equilibrium, or in another language, this is an instance of collective motion in the sense of Guillemin and Sternberg \cite{GS-book}.  The dynamics on $M$ is therefore integrable, and every motion takes place on a torus of dimension at most 2 (the rank of $SU(3)$). 

Moreover, since the Hamiltonian is a function of the distance, the distance for 2 point vortices is a conserved quantity and the generalized vortices cannot collide under the dynamics. In other words, the dynamics on $M^\circ$ is complete. 

A final observation is that, the set of orthogonal points (those at a distance of $\pi/2$) is an extremum of the Hamiltonian $h$, and consequently they necessarily form a group orbit of \emph{equilibria}. If the Hamiltonian is a strictly monotonic function of the distance, these will be the only equilibria in $M^\circ$.  Of course, if the Hamiltonian extends to $M$ then it will have critical points at the diagonal, which will therefore also consist of equilibria. In both cases, since the group orbits of equilibria are extremal for the Hamiltonian, they will be $SU(3)$-Lyapunov stable \cite{Mo97}, and even $SU(3)_\mu$-stable ($SU(3)_\mu$ is equal to $SU(3)$, or $U(2)$ or $\TT^2$ according to the configuration and the values of the $\Gamma_j$).

\begin{theorem} Consider the $SU(3)$ action on two vortices on $\CP ^2$, where every motion is a relative equilibrium. For any non-zero values of $\Gamma_1,\Gamma_2$, the space of allowed velocities is of dimension at most 1 according to the configuration type, as follows,
\begin{equation}
\dim R_0 = \begin{cases}
 0 & \text{if equal or orthogonal} \\
 1 & \text{otherwise}.
\end{cases}
\end{equation}
\end{theorem}

\begin{proof}
Table\,\ref{table:R0 for N=2} lists the space $R_0$ for every configuration, generic and otherwise, on $\CP ^2\times\CP ^2$.
\end{proof}

\begin{table}
\centering
\begin{tabular}{cc|c|}
	\cline{1-3}
\multicolumn{1}{ |c  }{\multirow{3}{*}{$\Gamma_1=\Gamma_2$} } &
\multicolumn{1}{ |c| }{equal: $G_x=U(2)=G_\mu$} & $R_0\simeq(\u(2)/\u(2))^{U(2)}=\{0\}$    \\[4pt] 
	\cline{2-3}
\multicolumn{1}{ |c  }{}                        &
\multicolumn{1}{ |c| }{orthogonal: $G_x=\TT^2$, $G_\mu=$U(2)} & $R_0\simeq(\u(2)/\tt^2)^{\TT^2}=\{0\}$   \\[4pt] 
	\cline{2-3}
\multicolumn{1}{ |c  }{}                        &
\multicolumn{1}{ |c| }{generic: $G_x=U(1)$, $G_\mu=\TT^2$} & $R_0\simeq(\tt^2/\u(1))^{U(1)}=\mathbb{R}$   \\[4pt] 
	\cline{1-3}
\multicolumn{1}{ |c  }{\multirow{3}{*}{$\Gamma_1=-\Gamma_2$} } &
\multicolumn{1}{ |c| }{equal: $G_x=U(2)$, $G_\mu=SU(3)$} & $R_0\simeq(\mathfrak{su}(3)/\u(2))^{U(2)}=\{0\}$     \\[4pt] 
	\cline{2-3}
\multicolumn{1}{ |c  }{}                        &
\multicolumn{1}{ |c| }{orthogonal: $G_x=G_\mu=\TT^2$} & $R_0\simeq(\tt^2/\tt^2)^{\TT^2}=\{0\}$   \\[4pt] 
	\cline{2-3}
\multicolumn{1}{ |c  }{}                        &
\multicolumn{1}{ |c| }{generic: $G_x=U(1)$, $G_\mu=\TT^2$} & $R_0\simeq(\tt^2/\u(1))^{U(1)}=\mathbb{R}$  \\[4pt] 
	\cline{1-3}
\multicolumn{1}{ |c  }{\multirow{3}{*}{Otherwise} } &
\multicolumn{1}{ |c| }{equal: $G_x=G_\mu=U(2)$} & $R_0 = \{0\}$   \\[4pt] 
	\cline{2-3}
\multicolumn{1}{ |c  }{} 	&
\multicolumn{1}{ |c| }{orthogonal: $G_x=G_\mu=\TT^2$} & $R_0 = \{0\}$   \\[4pt] 
	\cline{2-3}
\multicolumn{1}{ |c  }{} 	&
\multicolumn{1}{ |c| }{generic: $G_x=U(1)$, $G_\mu=\TT^2$} & $R_0\simeq(\tt^2/\u(1))^{U(1)}=\mathbb{R}$   \\[4pt] 
	\cline{1-3}
\end{tabular}

\medskip
\begin{minipage}{0.9\textwidth}
\caption{The allowed velocity spaces for relative equilibria for 2 vortices on $\CP^2$ (allowing for collisions).}
\label{table:R0 for N=2}
\end{minipage}
\end{table}

\subsection{Dynamics for 3 generalized point vortices}
Now let $M=\CP^2\times\CP^2\times\CP^2$, with vortex strengths $\Gamma_j$, and let $M^\circ$ be $M$ with the large diagonal removed (that is, omitting all collisions).

As discussed above (Theorem\,\ref{thm:reduction N=3}), each regular reduced space $M_\mu$ is diffeomorphic to a 2-sphere.  The reduced dynamics thereon will be Hamiltonian, with reduced Hamiltonian function $H_\mu$.   

There are some obvious conclusions to make: if $M_\mu$ is a sphere then there are at least 2 relative equilibria with that value $\mu$ of $\J$. If on the other hand, $M_\mu$ is a pointed sphere, there are also at least 2 relative equilibria, one of which must lie at the singular point.  Moreover, if there are just the two critical points on $M_\mu$, both relative equilibria are extremal and hence $G_\mu$-Lyapunov stable \cite{Mo97}.  If there are more than this minimum number of critical points, then some will be saddle points and hence unstable. 

Finally, if $M_\mu$ is a point, then it is a relative equilibrium, and trivially extremal, and hence $G_\mu$-Lyapunov stable.  

More interesting is the allowed velocities of the relative equilibria. In Table\,\ref{table:R0 for N=3} we assume the $\Gamma_j$ are generic, so the polytope is one of the 8 forms A, B,\dots, G, H described in \cite[Section 4]{MS19a} (see also Figure\,\ref{fig:polytopes+reduction} above).

\begin{table}
\centering
\begin{tabular}{|c|c|c|}
\multicolumn{3}{ c }{$\mu\in$ Wall:  $G_\mu=U(2)$}\\
\hline
triple point & $G_x=U(2)$ & $R_0\simeq(\u(2)/\u(2))^{U(2)}=\{0\}$    \\[4pt]
\hline
other vertices & $G_x=U(1)$ & $R_0\simeq(\u(2)/\u(1))^{U(1)} =\R^3$  \\[4pt] 
\hline
generic & $G_x=\mathbf{1}$ & $R_0\simeq\u(2) =\R^4$  \\[4pt] 
\hline
\end{tabular}

\bigskip

\begin{tabular}{|c|c|c|}
\multicolumn{3}{ c }{$\mu\not\in$ Wall: $G_\mu=\TT^2$}\\
\hline
double point & $G_x=U(1)$& $R_0\simeq(\tt^2/\u(1))^{U(1)}=\mathbb{R}$   \\[4pt] 
\hline
double+orthogonal & $G_x=\TT^2$& $R_0\simeq(\tt^2/\tt^2)^{\TT^2}=\{0\}$   \\[4pt] 
\hline
distinct coplanar & $G_x=U(1)$ & $R_0\simeq(\tt^2/\u(1))^{U(1)}=\mathbb{R}$   \\[4pt] 
\hline
totally orthogonal & $G_x=\TT^2$ & $R_0\simeq(\tt^2/\tt^2)^{\TT^2}=\{0\}$   \\[4pt]
\hline
semi-orthogonal & $G_x=U(1)$ & $R_0\simeq(\tt^2/\u(1))^{U(1)}=\mathbb{R}$   \\[4pt] 
\hline
generic  & $G_x=\mathbf{1}$ & $R_0\simeq\tt^2=\R^2$  \\[4pt] 
\hline
\end{tabular}

\medskip
\begin{minipage}{0.9\textwidth}
\caption{The allowed velocity spaces for relative equilibria for 3 vortices on $\CP^2$  (allowing for collisions), and for generic $\Gamma_j$. See text for explanations.}
\label{table:R0 for N=3}
\end{minipage}
\end{table}

\paragraph{Special configurations}
Consider the subgroup $\TT^2\subset SU(3)$ of diagonal matrices. The fixed points of $\TT^2$ in $\CP^2$ are $e_1,e_2$ and $e_3$, and therefore in $M$ the fixed points are the 27 points $(e_i,e_j,e_k)\in M$ for any $i,j,k\in\{1,2,3\}$.  Since these are isolated,  $\Fix(\TT,T_mM)=0$ and they are all necessarily equilibria for any $SU(3)$-invariant Hamiltonian.
If the Hamiltonian does not extend to allow collisions then the only ones of these allowed are the 6 totally orthogonal configurations (which map to the vertex $b$ under the orbit momentum map). 

Consider now the semi-orthogonal configurations, those where one of the points is orthogonal to the other two.  These points have stabilizer conjugate to $U(1)$, and in particular are all equivalent under the group action to points in $\Ma\simeq S^2\times S^2$, and are therefore necessarily relative equilibria.     

Finally, consider the coplanar configurations, corresponding to points in $\Mb$.  Now $\Mb\simeq S^2\times S^2\times S^2$, and the system reduces to that of three point vortices on the sphere. See for example \cite{PM98,LMR} for discussions of this system.

\paragraph{Identical vortices}  In this special case where $\Gamma_1=\Gamma_2=\Gamma_3$, there is a further symmetry of the system given by permutations of the point vortices.  Thus the full symmetry group becomes $G=SU(3)\times S_3$.

Now let $D\in SU(3)$ be any element of order 3; that is, one satisfying $D^3=I$, $D\neq I$.  In $SO(3)$ any element of order 3, if not the identity, is a rotation by $2\pi/3$ about some axis, and all subgroups of order 3 are conjugate.  However, in $SU(3)$ there are different (non-conjugate) elements of interest:
\begin{eqnarray*}
D_1 &=& \diag[1,\, \e^{2\i\pi/3},\, \e^{-2\i\pi/3}], \\
D_2 &=& \diag[\e^{2\i\pi/9},\, \e^{2\i\pi/9},\,\e^{-4\i\pi/9}].
\end{eqnarray*}
(In both cases $D_j^3$ is a scalar matrix which therefore acts trivially on $\CP^2$.) Let $\sigma=(1\;2\;3)\in S_3$, and let $\Sigma_j$ be the subgroup of order 3 of $G$ generated by $(D_j,\sigma)$. Now $m=(m_1,m_2,m_3)\in \Fix(\Sigma_j,M)$ if and only if
$$m_2=D_jm_1,\quad \text{and}\quad m_3=D_j^2m_1.$$
It follows that $\Fix(\Sigma_j,M) \simeq\CP^2$, parametrized by say $m_1\in\CP^2$ (for each of $j=1,2$).  The normalizer of $\Sigma_1$ is $\TT\times A_3$, while that of $\Sigma_2$ is $U(2)\times A_3$ (here $A_3$ is the cyclic subgroup of $S_3$ generated by $\sigma$). There are therefore actions of $\TT$ and $U(2)$ on $\Fix(\Sigma_1,M)$ and $\Fix(\Sigma_2,M)$ respectively, and the momentum maps for these actions are 
$$\J_{\Sigma_2}([x:y:z]) = \Gamma_1\begin{pmatrix}
3|x|^2-1&3x\overline y&0\cr 3\overline{x}y&3|y|^2-1&0\cr 0&0&3|z|^2-1 
\end{pmatrix}\in\u(2)^*
$$
and $\J_{\Sigma_1}([x:y:z]) = \diag[3|x|^2-1,\,3|y|^2-1,\,3|z|^2-1]\in\tt^*$, 
where $m_1=[x:y:z]\in\CP^2$.  A standard argument (see for example \cite[\S3.2]{LMR}) then shows that for identical vortices,  every configuration in $\Fix(\Sigma_j,M)$ is a relative equilibrium (for $j=1,2$).

\begin{remark}\label{rmk:pairwise interaction}
In this section we have only used the $SU(3)$-invariance of the Hamiltonian, and not the `pairwise interaction' form of \eqref{eq:pairwise interaction}.  It would be interesting to know (in general, not just in this context) what differences there are between Hamiltonian dynamics based on pairwise interactions, and more general (symmetric) Hamiltonian systems.  In the study of molecular dynamics, for example, the interactions between the atoms is not assumed to be a pairwise interaction, and the potential may depend on more general shape information.  
\end{remark}

\small
\setlength{\parskip}{0pt}

\bigskip
\setlength{\parindent}{0pt}

JM: j.montaldi@manchester.ac.uk,\quad
AS: amna.shaddad@gmail.com
\medskip

School of Mathematics, 
University of Manchester,
Manchester M13 9PL, UK

\end{document}

%% file: selectedPolytopes.tex

{\begin{subfigure}{0.28\textwidth}
\begin{tikzpicture}[scale=0.4] 
\WeylChamber
	\draw[very thick,fill=Crimson] (0, 8.25) -- (3.572, 2.062) -- 
	 (2.598, 1.5) -- 	(1.299, 1.5) -- (0, 3.75) -- (0, 8.25);
	\draw[thick] (1.299, 6) -- (0, 3.75);
	\draw[thick] (2.598, 3.75) -- (1.3,1.5);
	\draw[thick] (3.248, 2.625) -- (2.598, 1.5);
	 \draw [fill] (0, 8.25) node[anchor=east]{$a$};
	 \draw [fill] (1.3, 1.5) node[anchor=east]{$b$};
	 \draw [fill] (1.299, 6) circle [fill,radius=0.08] node[anchor=south west]{$c_3$};
	 \draw [fill] (2.598, 3.75) circle [fill,radius=0.08] node[anchor=south west]{$c_2$};
	 \draw [fill] (3.248, 2.625) circle [fill,radius=0.08] node[anchor=south west]{$c_1$};
 \end{tikzpicture}
\caption{Polytope A}
\label{subfig:N=3 A}
\end{subfigure}}
\qquad
{\begin{subfigure}{0.28\textwidth}
\begin{tikzpicture}[scale=0.4] 
\WeylChamber
	\draw[very thick,fill=Crimson] (0,4.5) node[anchor=east]{$a$} -- 
	(2.598, 9) node[black,anchor=south]{$c_3$} --
	(3.897, 6.7505) node[black,anchor=west]{$b$} --
	(1.299, 2.25) node[black,anchor=north]{$c_2$} -- (0,4.5);
	\draw[thick] (1.299, 6.75) -- (3.897, 6.7505);
	\draw [fill] (1.299, 6.75) circle [fill,radius=0.1] node[black,anchor=east]{$c_1$};
 \end{tikzpicture}
\caption{Polytope E}
\label{subfig:N=3 E}
\end{subfigure}}
\qquad
{\begin{subfigure}{0.28\textwidth}
\begin{tikzpicture}[scale=0.4] 
\WeylChamber
	\draw[very thick,fill=Crimson] (0,5.1) node[anchor=east]{$a$} -- 
	(2.078, 8.73) node[black,anchor=south]{$c_3$} --
	(4.676, 4.2) node[black,anchor=west]{$b$} --
	(3.377, 1.95) -- (2.208, 1.276) --
	(1.819, 1.95) -- (.5196, 4.22) -- (0,5.1);
	\draw[thick] (.5196, 4.22) -- (4.676, 4.2);
	\draw[thick] (3.377, 1.95) -- (1.819, 1.95);
	\draw (.5196, 4.22) circle [fill,radius=0.08] node[black,anchor=north]{$c_1\;$};
	\draw (1.819, 1.95) circle [fill,radius=0.08] node[black,anchor=east]{$c_2$};

 \end{tikzpicture}
\caption{Polytope H}
\label{subfig:N=3 H}
\end{subfigure}}
	
\medskip

%% file: PolytopesA-H+reduction.tex

\centering

{\begin{subfigure}{0.24\textwidth}
\begin{tikzpicture}[scale=0.35] 
\WeylChamber
	\fill[fill=Salmon] (0, 8.25) -- (3.572, 2.062) -- 
	 (2.598, 1.5) -- (1.299, 1.5) -- (0, 3.75) -- (0, 8.25);
	\draw[ultra thick,dashed,blue] (3.572, 2.062) -- (2.598, 1.5) -- 	(1.299, 1.5) -- (0, 3.75) -- (0, 8.25);
	\draw[thick] (1.299, 6) -- (0, 3.75);
	\draw[thick] (2.598, 3.75) -- (1.3,1.5);
	\draw[thick] (3.248, 2.625) -- (2.598, 1.5);
	 \draw (0, 8.25) node[anchor=east]{$a$};
	 \draw (1.3, 1.5) node[anchor=east]{$b$};
	 \fill (1.299, 6) circle [fill,radius=0.15] node[anchor=south west]{$c_3$};
	 \fill (2.598, 3.75) circle [fill,radius=0.15] node[anchor=south west]{$c_2$};
	 \fill (3.248, 2.625) circle [fill,radius=0.15] node[anchor=south west]{$c_1$};
 \end{tikzpicture}
\caption{Polytope A}
\label{subfig:N=3 A}
\end{subfigure}}
{\begin{subfigure}{0.24\textwidth}
\begin{tikzpicture}[scale=0.35] 
\WeylChamber
	\fill[fill=Salmon] (0,8.25) node[anchor=east]{$a$} -- 
	(1.3,6)  --  (1.948, 4.875)  --
	(3.248, 2.625) node[black,anchor=west]{$c_1$} --
	(0.6495, 2.625) node[black,anchor=north]{$b$} --
	(0,3.75) -- (0,8.25);
	\draw [thick] (1.948, 4.875) -- (0.6495, 2.625);
	\draw [thick] (1.3,6) -- (0,3.75);
	\draw [ultra thick,dashed,blue] (0,8.25) -- (0,3.75) -- (0.6495, 2.625) -- (3.248, 2.625) ;
	\fill (1.948, 4.875) circle [fill,radius=0.15]  node[anchor=south west]{$c_2$};
	\fill (1.3,6) circle [fill,radius=0.15] node[black,anchor=south west]{$c_3$} ;

 \end{tikzpicture}
\caption{Polytope B}
\label{subfig:N=3 B}
\end{subfigure}}
{\begin{subfigure}{0.24\textwidth}
\begin{tikzpicture}[scale=0.35] 
\WeylChamber
	\fill [fill=Salmon] (0, 6.75)  -- 
	(1.3, 9) node[black,anchor=west]{$c_3$} --
	(3.248, 5.625) node[black,anchor=west]{$b$} --
	(1.948, 3.375) node[black,anchor=north]{$c_2$} --
	(0.6495, 5.625) -- (0,6.75);
	\draw [thick] (3.248, 5.625) -- (0.6495, 5.625);
	\draw [ultra thick,dashed,blue]  (1.3, 9) -- (3.248, 5.625) -- (1.948, 3.375)  ;
	\fill (0.6495, 5.625) circle [radius=0.15] node[black,anchor=north]{$c_1$} ;
	\fill (0, 6.75) circle [radius=0.15] node[anchor=east]{$a$} ;
 \end{tikzpicture}
\caption{Polytope C}
\label{subfig:N=3 C}
\end{subfigure}}
{\begin{subfigure}{0.24\textwidth}
\begin{tikzpicture}[scale=0.35] 
\WeylChamber
	\fill[fill=Salmon] (0,1.5) node[anchor=east]{$a$} -- 
	(1.299, 3.751) -- (2.598, 6) -- 
	(3.897, 8.25) node[black,anchor=west]{$c_1$} --
	(1.299, 8.25) node[black,anchor=south]{$b$} --
	(0, 6) -- (0,1.5);
	\fill (1.299, 3.751) circle [fill,radius=0.15] node[black,anchor=west]{$c_2$};
	\fill (2.598, 6) circle [fill,radius=0.15] node[black,anchor=west]{$c_3$};
	\draw[thick] (2.598, 6) -- (1.299, 8.25);
	\draw[thick] (1.299, 3.751) -- (0,6); 
	\draw[ultra thick,dashed,blue] (0,1.5) -- (0,6) -- (1.299, 8.25) -- (3.897, 8.25);
 \end{tikzpicture}
\caption{Polytope D} 
\end{subfigure}}

\bigskip

{\begin{subfigure}{0.24\textwidth}
\begin{tikzpicture}[scale=0.35] 
\WeylChamber
	\fill[fill=Salmon] (0,4.5) -- 
	(2.598, 9) node [anchor=south]{$c_3$} --
	(3.897, 6.7505) node [anchor=west]{$b$} --
	(1.299, 2.25) node [anchor=north]{$c_2$} -- (0,4.5);
	\draw[ultra thick,dashed,blue] (2.598, 9)  -- 	(3.897, 6.7505)  --(1.299, 2.25);
	\draw[thick] (1.299, 6.75) -- (3.897, 6.7505);
	\fill (0,4.5)  circle [radius=0.15] node[anchor=east]{$a$};	
	\fill (1.299, 6.75) circle [radius=0.15] node[anchor=east]{$c_1$};
	\fill (0,4.5) circle [radius=0.15] ;
 \end{tikzpicture}
\caption{Polytope E}
\label{subfig:N=3 E}
\end{subfigure}}
{\begin{subfigure}{0.24\textwidth}
\begin{tikzpicture}[scale=0.35] 
\WeylChamber
	\fill [fill=Salmon] (0,3)  -- 
	(3.118, 8.4) node[black,anchor=south]{$c_3$} --
	(4.936, 5.25) node[black,anchor=west]{$b$} -- 
	(2.857, 1.65) -- (1.299, .75) -- 
	(.7794, 1.65)  -- (0,3);
	\draw[thick] (1.299, 5.25) -- (4.936, 5.25);
	\draw[thick] (.7794, 1.65) -- (2.857, 1.65);
	\draw[ultra thick,dashed,blue]  (3.118, 8.4)  -- (4.936, 5.25) -- (2.857, 1.65) -- (1.299, .75);
	\fill (0,3) circle [radius=0.15] node[anchor=east]{$a$}; 
	\fill (.7794, 1.65) circle [radius=0.15] node[black,anchor=east]{$c_2$};
	\fill (1.299, 5.250) circle [radius=0.15] node[black,anchor=east]{$c_1$};
\end{tikzpicture}
\caption{Polytope F}
\label{subfig:N=3 F}
\end{subfigure}}
{\begin{subfigure}{0.24\textwidth}
\begin{tikzpicture}[scale=0.35] 
\WeylChamber
	\fill [fill=Salmon] (0,1.5) node[anchor=east]{$a$} -- 
	(.7794, 2.85) circle [fill,radius=0.08] node[black,anchor=east]{$c_2$} --
	(1.819, 4.65) circle [fill,radius=0.08] node[black,anchor=east]{$c_1$} -- 
	(3.897, 8.25) node[black,anchor=south]{$c_3$} --
	(5.975, 4.65) node[black,anchor=west]{$b$} -- 
	(4.936, 2.850) -- (0.6494, 0.375) -- 
        (0,1.5);
	\draw[thick] (1.819, 4.65) -- (5.975, 4.65);
	\draw[thick] (.7794, 2.85) -- (4.936, 2.8);
	\draw[ultra thick,dashed,blue] (3.897, 8.25)  -- (5.975, 4.65)  -- (4.936, 2.850) -- (0.6494, 0.375) ;
	\fill (0,1.5) circle [radius=0.15] node [anchor=east]{$a$} ;
	\fill (.7794, 2.85) circle [radius=0.15] node [anchor=east]{$c_2$} ;
	\fill (1.819, 4.65) circle [radius=0.15] node [anchor=east]{$c_1$} ;
\end{tikzpicture}
\caption{Polytope G}
\label{subfig:N=3 G}
\end{subfigure}}
{\begin{subfigure}{0.24\textwidth}
\begin{tikzpicture}[scale=0.35] 
\WeylChamber
	\fill[fill=Salmon] (0,5.1) -- 
	(2.078, 8.73) node[black,anchor=south]{$c_3$} --
	(4.676, 4.2) node[black,anchor=west]{$b$} --
	(3.377, 1.95) -- (2.208, 1.276) --
	(1.819, 1.95) -- (.5196, 4.22) -- (0,5.1);
	\draw[ultra thick,dashed,blue] (2.078, 8.73)  --
	(4.676, 4.2)  -- 	(3.377, 1.95) -- (2.208, 1.276) ;
	\draw[thick] (.5196, 4.22) -- (4.676, 4.2);
	\draw[thick] (3.377, 1.95) -- (1.819, 1.95);
	\fill (0,5.1) circle [radius=0.15] node[anchor=east]{$a$};
	\fill (.5196, 4.22) circle [radius=0.15] node[anchor=north]{$c_1\;$};
	\fill (1.819, 1.95) circle [radius=0.15] node[anchor=east]{$c_2$};
 \end{tikzpicture}
\caption{Polytope H}
\label{subfig:N=3 H}
\end{subfigure}}

\bigskip